\documentclass[12pt]{iopart}

\usepackage{times}
\usepackage{gensymb}
\expandafter\let\csname equation*\endcsname\relax
\expandafter\let\csname endequation*\endcsname\relaxr
\usepackage{amsmath}
\usepackage{graphicx}
\usepackage{pdflscape}
\usepackage{csvsimple}
\usepackage{booktabs}
\usepackage{threeparttable}
\usepackage[T1]{fontenc}
\usepackage{textcomp}
\usepackage{makecell}
\usepackage{adjustbox}
\usepackage{lineno}
\usepackage{nomencl}

\makenomenclature

\usepackage{etoolbox}
\renewcommand\nomgroup[1]{%
  \item[\bfseries
  \ifstrequal{#1}{C}{CO\textsubscript{2}}{%
    \ifstrequal{#1}{G}{Other GHG emissions}{%
      \ifstrequal{#1}{A}{Aerosol and pollutants}{%
        \ifstrequal{#1}{O}{Ozone}{%
          \ifstrequal{#1}{T}{Other}{%
            \ifstrequal{#1}{S}{Surface albedo}{%
              \ifstrequal{#1}{m}{Climate}
            }
          }
        }
      }
    }
  }%
]}

\usepackage{chapterbib}
\usepackage{hyperref}

\hypersetup{
    colorlinks=true,
    linkcolor=blue,
    citecolor=blue,
    urlcolor=blue
}

\usepackage{cite}

\usepackage[capitalise]{cleveref}

\usepackage{tikz}

\newcommand*\circled[1]{\tikz[baseline=(char.base)]{%
            \node[shape=circle,draw,inner sep=2pt] (char) {#1};}}

\usepackage{enumitem}

\begin{document}
% \linenumbers

\title{Source attributions of radiative forcing by regions, sectors, and climate forcers}

\author{Xuanming Su$^{1\ast}$, Kaoru Tachiiri$^{1,2}$, Katsumasa Tanaka$^{2,3}$, Michio Watanabe$^1$ \& Michio Kawamiya$^{1}$}

\address{$^{1}$Research Institute for Global Change / Research Center for Environmental Modeling and Application / Earth System Model Development and Application Group, Japan Agency for Marine-Earth Science and Technology (JAMSTEC), Yokohama, Japan}

\address{$^{2}$Center for Global Environmental Research, National Institute for Environmental Studies (NIES), Tsukuba, Japan}

\address{$^{3}$Laboratoire des Sciences du Climat et de l’Environnement (LSCE), Commissariat à l’énergie atomique et aux énergies alternatives (CEA), Gif-sur-Yvette, France}

% \address{$^{4}$Research Institute for Global Change/Research Center for Environmental Modeling and Application/Climate Model Development and Application Group, Japan Agency for Marine-Earth Science and Technology}

% \address{$^{5}$Research Institute for Global Change/Research Center for Environmental Modeling and Application, Japan Agency for Marine-Earth Science and Technology}

\ead{$^{*}$suxuanming@jamstec.go.jp}
\vspace{10pt}

% \begin{bibunit}[iopart-num]

\begin{abstract}
It is important to understand how the emissions of different regions, sectors, or climate forcers play a role on pathways toward the Paris Agreement temperature targets. There are however methodological challenges for attributing individual contributions due to complexities associated with a variety of climate forcers affecting the climate system on different spatial and temporal scales. Here, we use the latest historical and future emissions data for a comprehensive set of climate forcers as well as land-use datasets and apply the normalized marginal approach to quantify the forcing contributions of regions, sectors and forcing agents toward the 2 \textdegree C and 1.5 \textdegree C targets. We show that most of the worldwide regions and sectors need to maintain forcing levels not higher than present levels to attain the 1.5 \textdegree C target of the Paris Agreement, while slightly higher future forcing levels than present levels are allowed for the 2 \textdegree C target. Our results illustrate the importance of negative CO\textsubscript{2} emissions, which contribute -$0.75\pm 0.44$ Wm\textsuperscript{-2} and -$0.42\pm 0.27$ Wm\textsuperscript{-2} to the 2 \textdegree C and 1.5 \textdegree C targets. Less negative forcings, or more positive forcings are also identified for the land-use albedo for the 2 \textdegree C and 1.5 \textdegree C scenarios compared to existing studies.
\end{abstract}

\section{Introduction}

The Paris Agreement has set goals to limit the global average temperature increase to well below 2 \textdegree C and to pursue efforts to limit the temperature increase to 1.5 \textdegree C above the preindustrial level. The Paris Agreement goals can be translated into the required levels of greenhouse gas (GHG) emission reductions \cite{Rogelj2018, Tanaka2018, sr15, Tong2019, Tachiiri2019, Kawamiya2019} for practical implementation purposes, through calculations of the radiative forcings resulting from various emission sources. Attaining the radiative forcing of 2.6 Wm\textsuperscript{-2} and 1.9 Wm\textsuperscript{-2} are known to be largely consistent with achieving the 2 \textdegree C and 1.5 \textdegree C climate goals, respectively, at an approximately 66\% probability \cite{Meinshausen2009,ar5ch8,Rogelj2018,VanVuuren2018,Seneviratne2018,sr15}. Here we use the radiative forcing as a benchmark and assess how individual regions, sectors, and climate forcers can contribute to achieving the Paris Agreement temperature targets based on a latest set of historical and future emission data. Such source attribution cannot be done from emissions data alone because a variety of climate forcers affect the climate system on different temporal and spatial scales in nonlinear ways, requiring dedicated methodologies like the one presented here. Attributing forcing at the levels of regions, sectors, or climate forcers provides a basis for considering the principle of common but differentiated responsibilities contained in the 1992 United Nations Framework Convention on Climate Change (UNFCCC). 

We identified two issues associated with previous attribution methods. First, comprehensive regional and sectoral assessments should in principle consider a full suite of anthropogenic sources at the regional and sectoral levels, including GHGs, aerosols and pollutants, as well as land-use albedo. However, not all sources have been considered in previous studies. For example, aerosols and pollutants were not always examined\cite{Rive2008, denElzen2013}, or only a subset of aerosol species was considered (such as sulfate aerosols\cite{Matthews2014}). Also, land-use albedo were sometimes not included\cite{Rive2008, denElzen2013, Matthews2014}. This may be due to the lack of datasets or difficulties in representing regional or sectoral forcings, but the latest datasets provide opportunities to consider a more comprehensive set of GHGs and related agents. Second, among various methods proposed, only the marginal and time-sliced methods are considered to be useful based on a satisfaction test of eight essential criteria\cite{IPCC2002, Trudinger2005}. In principle, an attribution method needs to ensure additivity regarding regions and time. Implementing a non-additive method like the residual method used in \cite{Skeie2017} may introduce bias to the outcome for regions with high and low emissions. Some of the methods yield non-zero radiative forcing when a region's concentration has become zero. On the other hand, the two recommended methods are computationally expensive, especially when various sources of uncertainties are considered.

We apply the most up-to-date emissions and land-use datasets (\cref*{tbl_hist,tbl_sce,tbl_luc,fig:ghg,fig:aero1,fig:aero2,fig:lc}), which resolve regions and sectors and contain all pertinent forcing sources, including GHGs, aerosols and pollutants, and land-use albedo, as well as their associated uncertainties. Particularly we consider a range of future emissions trajectories with various socioeconomic backgrounds and climate mitigation levels\cite{Fujimori2018,Gidden2019}. We combine the normalized marginal method, which is computationally less expensive than the time-sliced method\cite{IPCC2002}, with a simple climate model - the Simple Climate Model for Optimization version 2 (SCM4OPT v2.0) \cite{Su2017, Su2018}. SCM4OPT v2.0 is designed to be lightweight and suitable for performing a large number of simulations required for our study exploring uncertainties, while resolving diverse characteristics of forcing agents considered. Furthermore, based on the premise of previous studies\cite{Rive2008, Hohne2011, Li2016}, we make the attribution analysis more comprehensive by considering historical and future emissions and providing perspectives from regions, sectors, and climate forcers. We consider two types of uncertainties, including those due to our lack of knowledge regarding historical emissions and future projections (emission uncertainties), and those due to low confidence in understanding the climate system (climate uncertainties). With these methodological advances, we quantify the forcing contributions of regions, sectors, and climate forcers toward the Paris Agreement temperature goals.

\section{Methods}

We compile the emissions and land cover datasets at regional and sectoral levels and implement them to SCM4OPT v2.0 to calculate the marginal forcing effects of forcing agents at regional and sectoral levels. The relative forcing contribution of the forcing agents at regional and sectoral levels can be distinguished based on the fraction accounting for their marginal effects in total marginal effects caused by the forcing agent. The forcing contributions of the forcing agents at regional and sectoral levels can be therefore attributed. We sum up the associated individual forcings to obtain the radiative forcings resulting from regional and sectoral sources.

\subsection{Emission data and uncertainties} We used historical and future emissions and land cover datasets with both regional and sectoral details. The available datasets are shown in \cref*{tbl_hist,tbl_sce,tbl_luc,fig:ghg,fig:aero1,fig:aero2,fig:lc} \cite{Lamarque2009,Gutschow2016,Aardenne2018,Hoesly2018, Fujimori2018,Gidden2019}. We designated historical sources as those originating from 1850 to 2016, and future projections by 2100 are grouped by the forcing level at 1.9 Wm\textsuperscript{-2} and 2.6 Wm\textsuperscript{-2}, regardless of the underlying socioeconomic development or technological assumptions. We also included other scenarios with relatively lower possibilities to achieve the 2 \textdegree C and 1.5 \textdegree C targets, namely, forcing levels of 3.4 Wm\textsuperscript{-2}, 4.5 Wm\textsuperscript{-2}, 6.0 Wm\textsuperscript{-2}, 7.0 Wm\textsuperscript{-2} and 8.5 Wm\textsuperscript{-2} (\cref*{tbl_sce}). Thus, a broad range of forcings can be examined for future climate change. For the future emission datasets, 25 of them are obtained from the Asia-Pacific Integrated Model/Computable General Equilibrium (AIM/CGE) model\cite{Fujimori2018}, and the remaining nine are from the Integrated Assessment Modeling Consortium (IAMC)\cite{Gidden2019} (\cref*{tbl_sce}). We divided the world into eleven regions, including 1) China (CHN), 2) India (IND), 3) Japan (JPN), 4) Russia (RUS), 5) the United States of America (USA), 6) sub-Saharan Africa (AFR), 7) Europe (EUR), 8) Latin America and the Caribbean (LAM), 9) the Middle East and North Africa (MEA), 10) other areas in Asia (OAS) and 11) the rest of the world (ROW) (\cref*{tbl_region}). For each region, twelve emitting sectors \cite{Lamarque2009, Fujimori2018, Gidden2019} were assessed, namely, 1) agriculture, 2) agricultural waste burning, 3) domestic and commercial housing, 4) energy, 5) industry, 6) industrial solvents, 7) surface transportation, 8) waste treatment, 9) open forest burning, 10) open grassland burning, 11) aviation and 12) international shipping (\cref*{tbl_sector}). In addition to the twelve sectors above, 13) land-use CO\textsubscript{2} emissions and 14) negative CO\textsubscript{2} emissions, namely, through carbon capture and storage (CCS) and bioenergy with CCS (BECCS), were separately considered. We compiled the emissions from the available datasets into scenario-, region- and sector-specific emissions and used $E_{n,r,s,e}(t)$ to denote scenario ($n$)-, region ($r$)- and sector ($s$)-specific emissions ($e$, refer to the emission species) over time $t$. The emissions of the same forcing target originating from different Shared Socioeconomic Pathways (SSPs) and integrated assessment models (IAMs) were treated as emission uncertainties. For example, we iteratively simulated the 1.9 Wm\textsuperscript{-2} forcing scenario by using dataset from the AIM/CGE (SSP1-1.9 and SSP2-1.9) and the IAMC (SSP1-1.9), as reported in \cref*{tbl_sce}.

\subsection{Climate model and uncertainties}

We used the simple climate model SCM4OPT v2.0 to generate the outputs for our analysis. The current model has been updated from the precedent in the following four respects: First, we adopted the ocean carbon cycle of Hector v1.0\cite{Hartin2015} and applied the Diffusion Ocean Energy balance CLIMate (DOECLIM) model \cite{Kriegler2005,Tanaka2007,Wong2017} to calculate the temperature change. We calibrated the carbon cycle and temperature modules based on 26 coupled atmosphere-ocean general circulation models (AOGCMs) with outputs for the carbon cycle in the Coupled Model Intercomparison Project, Phase 5 (CMIP5) (\cref*{tbl_clims}). Second, parameters associated with CH\textsubscript{4}, N\textsubscript{2}O and halogenated gases (a total of 37 gases, see \cref*{tbl_driver}) were tuned against the atmospheric lifetimes and radiative efficiencies in the IPCC Fifth Assessment Report (AR5) \cite{ar5annexii}. Third, we employed the simple global parameterizations described in OSCAR v2.2 \cite{Gasser2016} to estimate the radiative forcings resulting from aerosols and pollutants. The radiative forcing of short-lived climate forcers depends on the geographical location of emissions. The spatial distribution of the radiative forcing of short-lived species is different from that of long-lived species\cite{Sand2016,Tanaka2019}. However, these two effects are not considered in our analysis. Fourth, we adopted a simple parameterization scheme \cite{Gasser2016} to calculate the land-use albedo (see \cref*{eq:flcc} in the supplementary materials). The equations for the climate model are listed in the supplementary materials. 

We performed a robustness test over the historical period by using historical emission datasets as input and considering the climate uncertainties that were applied in this analysis. The outputs from our model are consistent with those of other models or statistical records (\cref*{fig:rfc_ghg,fig:rfc_aero,fig:rfc_other,fig:rfc_natr,fig:rfc_tot,fig:scm_tatm}). Furthermore, the likelihoods of meeting the 2\textdegree C and 1.5 \textdegree C targets of each of the forcing scenarios obtained from our model agree largely with the corresponding IPCC ranges (\cref*{fig:densi}) to limit global warming to 2\textdegree C with at least 66\% probability and 1.5 \textdegree C with 50\%\cite{sr15}.

\subsection{Calculation of the regional and sectoral forcings}

We utilized and expanded the normalized marginal method presented in ref \cite{IPCC2002, Trudinger2005, Li2016} to conduct our analysis. The relative forcing contribution of emission $E_{n,r,s,e}$ (Column 2 in \cref*{tbl_driver}) to the associated radiative forcing $f$ (Column 4 in \cref*{tbl_driver}), which is defined as $\alpha^{f}_{n,r,s,e}$, is proportional to the marginal effect of $E_{n,r,s,e}$ causing the radiative forcing $f$ (see \cref*{fig:margin}). To calculate $\alpha^{f}_{n,r,s,e}$, for each $E_{n,r,s,e}$, we performed two simulations, i.e., 1) one simulation with all emissions included in the simulation as input, to calculate the associated radiative forcing termed $F^{all,f}_{n,r,s,e}$, and 2) another simulation with the emission $e$ reduced by $E_{n,r,s,e}\cdot\epsilon$ ($\epsilon = 0.001$) over the evaluation period, that is 1850-2100, to obtain the corresponding radiative forcing named $F^{\epsilon,f}_{n,r,s,e}$. The relative contribution $\alpha^{f}_{n,r,s,e}$ is obtained by:

\begin{equation}
  \label{eq:marginal1}
\alpha^{f}_{n,r,s,e} = \frac{F^{all,f}_{n,r,s,e}-F^{\epsilon,f}_{n,r,s,e}}{\sum_{r,s,e}{\left(F^{all,f}_{n,r,s,e}-F^{\epsilon,f}_{n,r,s,e}\right)}} 
\end{equation}

Therefore, the radiative forcing $F^{f}_{n,r,s,e}$, which is resulting from $E_{n,r,s,e}$, is isolated by:

\begin{equation}
  \label{eq:marginal2}
F^{f}_{n,r,s,e} = {F^{all,f}_{n,r,s,e}} \cdot \alpha^{f}_{n,r,s,e} 
\end{equation}

To consider the relevant emission and climate uncertainties, we carried out 200 similar pairs of runs for each forcing-level-specific source $E_{n,r,s,e}$, with randomized scenarios at the same forcing level and randomized parameter sets for the climate system, and we call them as one experiment for $E_{n,r,s,e}$ (see the randomization sources of the scenarios (within each forcing level) and the climate system in \cref*{tbl_unc}). Here, the value of 200 has been tested to ensure that two decimal places of the precision level could be achieved for the mean forcing value of $F^{f}_{n,r,s,e}$ under different experiments. 

We obtain the individual forcing agents by summing the forcings induced by all available emissions sources. Therefore, a certain forcing agent is probably a mixed effect resulting from various emissions sources. On the other hand, a particular emission may result in different kinds of radiative forcings, as indicated in \cref*{tbl_driver}. For example, black carbon (BC) can cause BC forcing, BC on snow and indirect cloud effects. A total of $5.3\times 10^{6}$ runs ($7.6\times 10^{5}$ for each forcing level) were performed considering all forcing levels, regions, sectors and emissions. To derive the regional forcings, we applied the Monte Carlo approach (n = 20,000) to sum all $F^{f}_{n,r,s,e}$ values belonging to a given region. Here, the value of 20,000 for n was also tested to ensure the necessary precision for our analysis. The sectoral forcings are similarly obtained. An overview of all the iterations are contained in \cref*{tbl_unc}.

\subsection{The probability of exceeding 2\textdegree C or 1.5\textdegree C} For each experiment, 200 sample results were acquired. Here, we assumed that the obtained temperature increase $T$ over time $t$ followed a normal distribution, and the cumulative distribution function was defined as:

\begin{equation}
  \label{eq:cdf}
  F^{t}_{T}\left(\tau\right) = P^{t}\left(T\leq\tau\right)
\end{equation}

We used the exceedance of \cref*{eq:cdf} to obtain the probability of exceeding a specified climate target $\tau$:

\begin{equation}
  \label{eq:exceed}
  \overline{F}^{t}_{T}\left(\tau\right) = P^{t}\left(T > \tau\right) = 1 - F^{t}_{T}\left(\tau\right)
\end{equation}

Therefore, $\overline{F}^{t}_{T}\left(2 \right)$ indicates the probability of exceeding 2\textdegree C, and $\overline{F}^{t}_{T}\left(1.5 \right)$ gives the probability of exceeding 1.5\textdegree C. 
\section{Results}

\subsection{Regional attributions}
We performed our analysis based on the available existing scenarios, and the 2 \textdegree C or 1.5 \textdegree C results herein thus reflect the diagnosed compatible scenarios in terms of the 2 \textdegree C or 1.5 \textdegree C targets. The results reveal that the USA, China and the European Union (EU) are three major emitters, accounting for approximately 45\% of all the forcings under the historical, 2 \textdegree C and 1.5 \textdegree C scenarios (see \cref*{fig:rfc_reg}a). China's share increased from $12\pm 4$\% ($0.25\pm 0.09$ Wm\textsuperscript{-2}) by 2016 (cf. $10 \pm 4$\% for Chinese data (1750-2010) in ref \cite{Li2016} with similar methods) to 2 \textdegree C's $16\pm 3$\% ($0.41\pm 0.08$ Wm\textsuperscript{-2}) and 1.5 \textdegree C's $17\pm 4$\% ($0.29\pm 0.07$ Wm\textsuperscript{-2}), while the share of the EU declined, from the historical $15\pm 2$\% ($0.32\pm 0.04$ Wm\textsuperscript{-2}) level to the 2 \textdegree C level of $12\pm 2$\% ($0.31\pm 0.06$ Wm\textsuperscript{-2}) and the 1.5 \textdegree C level of $13\pm 3$\% ($0.23\pm 0.06$ Wm\textsuperscript{-2}) (for the forcing values, see \cref*{fig:sec_all}a\&\cref*{fig:reg_grp}). In contrast, the share of the USA exhibited no major changes, contributing to approximately 17\% of the total forcings under all three scenarios. However, the absolute values of the forcings varied, i.e., the 2 \textdegree C forcing attributed to the USA increased to $0.42\pm 0.07$ Wm\textsuperscript{-2} from the current forcing value of $0.36\pm 0.05$ Wm\textsuperscript{-2}, while the 1.5 \textdegree C forcing value declined to $0.29\pm 0.06$ Wm\textsuperscript{-2}. Latin America and the Caribbean (as one region) also exhibited a relatively high historical share with $12\pm 3$\%; however, the value substantially declined under both target scenarios.

CO\textsubscript{2}, including fuel CO\textsubscript{2}, land-use CO\textsubscript{2}, and negative CO\textsubscript{2}, if applicable, is the main contributor and varies across regions. Among them, China, the USA, and the Middle East and North Africa (as one region) exhibited the highest net growth forcings under the 2 \textdegree C scenario, with values of $0.08\pm 0.14$ Wm\textsuperscript{-2}, $0.07\pm 0.13$ Wm\textsuperscript{-2} and $0.07\pm 0.06$ Wm\textsuperscript{-2}, respectively. Under the 1.5 \textdegree C scenario, the CO\textsubscript{2} forcings in all regions decreased. The largest decline occurred in Latin America and the Caribbean, from the historical value of $0.20\pm 0.04$ Wm\textsuperscript{-2} to the 1.5 \textdegree C scenario value of $0.08\pm 0.05$ Wm\textsuperscript{-2}, which occurred due to the negative CO\textsubscript{2} emissions and the great decrease in land-use CO\textsubscript{2} emissions.

The non-CO\textsubscript{2} forcings described here refer to the forcings induced by sources other than CO\textsubscript{2}, and these forcings also play an important role in the historical period, although they are almost adequately controlled under the 2 \textdegree C and 1.5 \textdegree C scenarios (also shown in \cref*{fig:reg_grp}). Basically, most of the regions reveal net positive non-CO\textsubscript{2} forcings in the historical period. Particularly in regard to Latin America and the Caribbean, the relatively high net positive non-CO\textsubscript{2} forcing, combined with the relatively high land-use CO\textsubscript{2} forcing, contributes to a comparatively large forcing share in the historical period, although the fossil-fuel forcing is relatively smaller. It is worth noting that the regions with nearly zero-sum non-CO\textsubscript{2} forcings contribute considerable amounts of both positive and negative forcings, such as China and the rest of the world, in the historical period. For future non-CO\textsubscript{2} forcings, however, sub-Saharan Africa is found to exhibit a reasonable increase in net forcing due to its continuous development and industrialization and population growth, which requires more biomass for cooking and heating purposes, as well as changes in land cover \cite{Fujimori2018,Gidden2019}.

The total forcing, including the forcings that cannot be assigned to any region, increases to $2.6\pm 0.4$ Wm\textsuperscript{-2} under the 2 \textdegree C scenario but declines to $1.8\pm 0.4$ Wm\textsuperscript{-2} under the 1.5 \textdegree C scenario, which is lower than the current level of $2.2\pm 0.4$ Wm\textsuperscript{-2} (\cref*{fig:rfc_reg}b). All regional forcings indicate net warming effects, with positive forcing values. Forcing increases are encountered in most of the regions except in Russia, the EU, Latin America and the Caribbean, and the rest of the world under the 2 \textdegree C scenario, while the main increases still occur in two developing regions, namely, China and the Middle East and North Africa under the 1.5 \textdegree C scenario (\cref*{fig:sec_all}a\&\cref*{fig:reg_grp}). Here, the forcing increases in the Middle East and North Africa can mostly be attributed to fossil-fuel CO\textsubscript{2}, sulfate, cloud effects, and land-use albedo, probably due to industry and energy supply expansions as well as due to the expected reforestation in this area \cite{Fujimori2018}. The regional nonattributable forcings in the historical period also reveal warming effects. These forcings are later notably suppressed under both the 2 \textdegree C and 1.5 \textdegree C scenarios (\cref*{fig:rfc_reg}b), and they are mainly attributed to the control of ozone-depleting substances (ODSs) under the various scenario assumptions (\cref*{fig:ghg})\cite{Fujimori2018,Gidden2019}.

{\linespread{1}
\begin{figure}[ht]
  \includegraphics[width=0.75\textwidth]{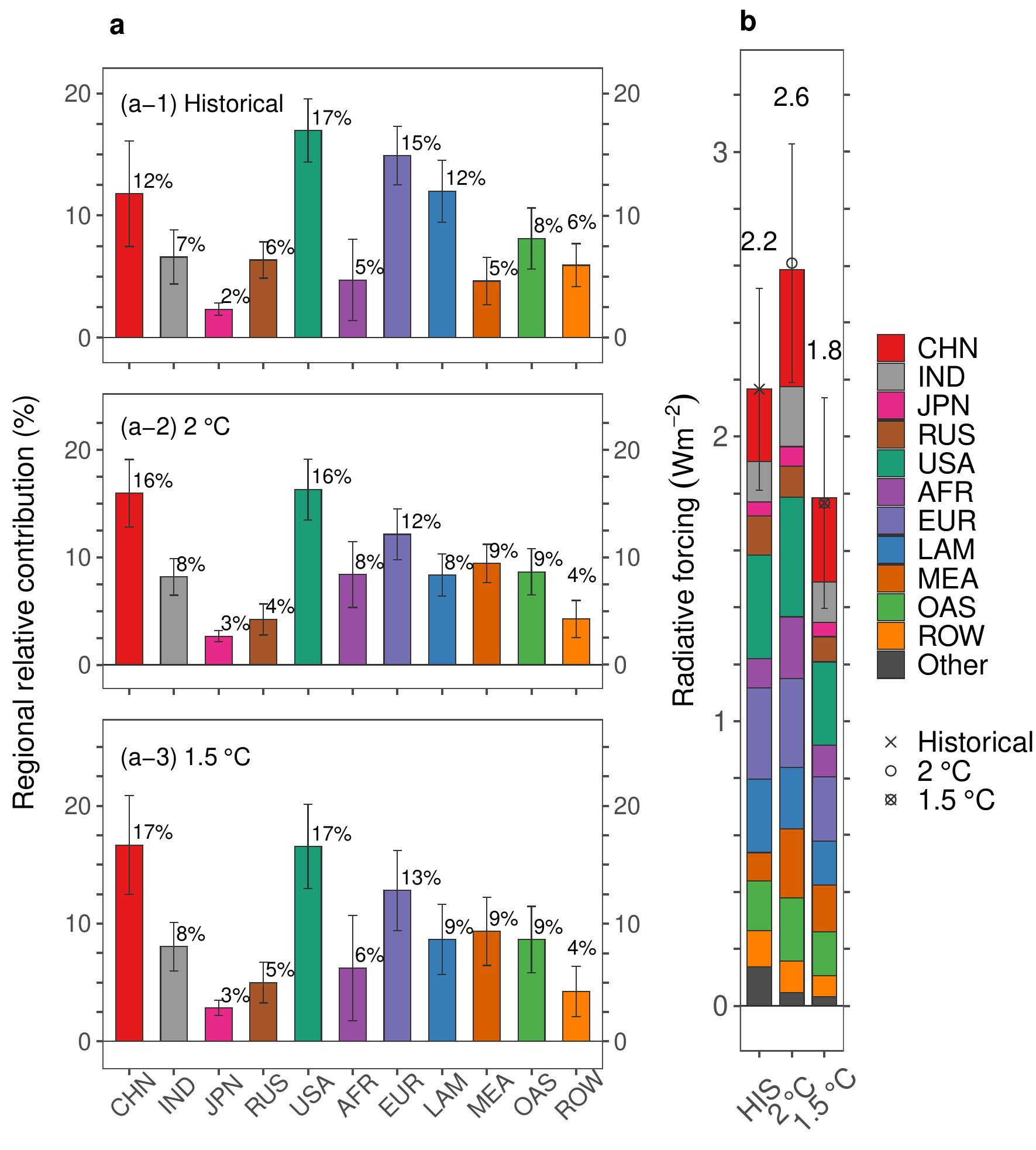}
  \centering
  \caption{\small{\textbf{Reginal contributions to climate change.} a, Regional relative contributions to climate change. The regional relative contributions are derived from the elementwise ratio of the regional forcings to the total forcings via the methods described in ref \protect\cite{Li2016}. Note that the sum of the mean percentages of all regions is not equal to 100\% since regional nonattributable forcings occur (see \cref*{fig:rfc_reg}b). b, The world total forcings are divided into regional forcings. The value on top of the bar indicates the mean value of the total radiative forcing, and the error bar indicates the associated uncertainty resulting from the world ensemble. The other forcings in Fig. b are the regional nonattributable climate forcers, including the international shipments of $0.01\pm 0.03$, $0.06\pm 0.02$, and $0.05\pm 0.03$ Wm\textsuperscript{-2} and part of the ozone-depleting substances (ODSs) of $0.23\pm 0.05$, $0.07\pm 0.03$, and $0.07\pm 0.03$ Wm\textsuperscript{-2} (for the regional nonattributable forcings, see \cref*{tbl_driver}), under the historical, 2 \textdegree C and 1.5 \textdegree C scenarios, as well as mineral dust (\cref*{fig:rfc_aero}) and the effects of solar irradiance and volcanic activity (\cref*{fig:rfc_natr}). The probabilities of reaching the 2 \textdegree C and 1.5 \textdegree C targets here are 56\% and 61\%, respectively (see \cref*{fig:rfc_prob}). All uncertainties are represented as one standard deviation. CHN, China; IND, India; JPN, Japan; RUS, Russia; USA, United States of America; AFR, sub-Saharan Africa; EUR, Europe; LAM, Latin America and the Caribbean; MEA, Middle East and North Africa; OAS, other Asian countries; ROW, the rest of the world; Other, regional nonattributable forcings.}}
  \label{fig:rfc_reg}
\end{figure}
}

\subsection{Sectoral constituents}

The regional effects are further separated into their sectoral constituents to assess how future changes occur (\cref*{fig:sec_all}a). First, for the developed regions, relatively large increases are observed in both industrial and housing sectors under the 2 \textdegree C scenario. In regard to energy, the gross forcings related to the USA and EU are considerably high under the 2 \textdegree C scenario. However, if combined with the negative CO\textsubscript{2} emissions, the energy forcings decrease to $0.12\pm 0.11$ Wm\textsuperscript{-2} and $0.07\pm 0.08$ Wm\textsuperscript{-2} for the USA and EU, respectively, which are lower than the current levels. Second, among the developing regions, China's industry exhibits the most significant forcing increase, with a value of $0.14\pm 0.07$ Wm\textsuperscript{-2} under the 2 \textdegree C scenario. In addition to the industrial sector, prominent increases are found in the agricultural sector, such as in sub-Saharan Africa, and the forcing induced by the agricultural sector increases by $0.09\pm 0.03$ Wm\textsuperscript{-2} under the 2 \textdegree C scenario. In addition, the land-use CO\textsubscript{2} forcings are alleviated to varying degrees in all regions under the 2 \textdegree C scenario. Under the 1.5 \textdegree C scenario, most of the regions still demonstrate increased forcings in the industrial sector, while in the developed regions, the forcings in the industrial sector decrease. Furthermore, both the negative and land-use CO\textsubscript{2} emissions could result in extensive forcing abatement from the current levels in the developing regions under the 1.5 \textdegree C scenario.

Globally, in certain sectors, as shown in \cref*{fig:sec_all}b, the forcings still increase to certain levels under the 2 \textdegree C scenario, except for the land-use CO\textsubscript{2} and other sources that are responsible for the main emissions of aerosols and pollutants, such as waste treatment, agricultural waste burning, forest burning and grass burning. However, under the 1.5 \textdegree C scenario, except for the energy sector and land-use albedo, only a small amount of the forcings is found to increase in the major emitting sectors, such as domestic and commercial housing, industrial sector, aviation and international shipping. In addition, as also indicated in the analysis of the individual forcing agents below, the negative CO\textsubscript{2} emissions remove a considerable amount of forcings from the energy sector under the 1.5 \textdegree C scenario, and the net forcing level in the energy sector is lower than the current energy sector level. This result implies that to attain the 1.5 \textdegree C target, efforts need to be implemented to maintain the sectoral forcings below or equal to the current  levels.

{\linespread{1}
\begin{figure}[ht]
  \includegraphics[width=0.9\textwidth]{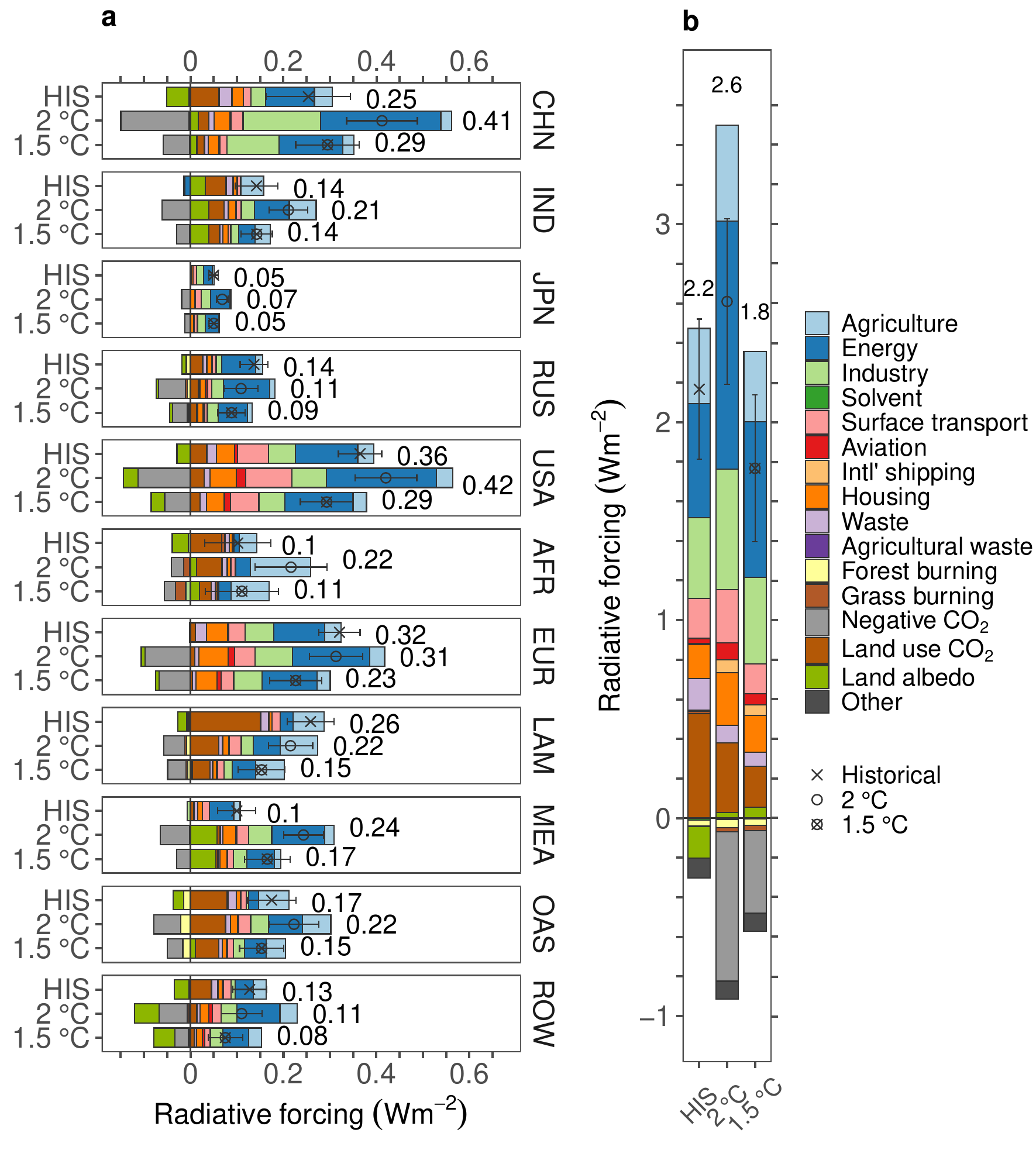}
  \centering
  \caption{\small{\textbf{Sectoral contributions to climate change.} a, Sectoral contributions of the eleven regions worldwide. b, The world total forcings are decomposed into sectoral forcings. The value on top of the bar indicates the mean value of the total radiative forcing, and the error bar indicates the associated uncertainty resulting from the world ensemble. The other forcings are the forcings induced by mineral dust (\cref*{fig:rfc_aero}), solar irradiance and volcanic activity (\cref*{fig:rfc_natr}). The probabilities of reaching the 2 \textdegree C and 1.5 \textdegree C targets here are 56\% and 61\%, respectively (see \cref*{fig:rfc_prob}). All uncertainties are represented as one standard deviation.}}
  \label{fig:sec_all}
\end{figure}
}

\subsection{Individual forcing agents}

\cref*{fig:rfc_grp}a shows the individual forcing agents for each sectoral source. Fossil-fuel CO\textsubscript{2} dominates the forcings in the housing, energy, industrial and transport sectors, particularly under the 2\textdegree C and 1.5 \textdegree C scenarios when the other GHGs and aerosols and pollutants are substantially removed (\cref*{fig:ghg,fig:aero1,fig:aero2}) and the resulting impacts are therefore greatly reduced.

Among them, first, the negative CO\textsubscript{2} emissions can eliminate considerable amounts of forcings. For example, -$0.75\pm 0.44$ Wm\textsuperscript{-2} and -$0.42\pm 0.27$ Wm\textsuperscript{-2} are attributed to the negative CO\textsubscript{2} emissions under the 2 \textdegree C and 1.5 \textdegree C scenarios, respectively, (\cref*{fig:rfc_grp}a). It is interesting to note that the reduced amount of the absolute forcing under the 2 \textdegree C scenario is even larger than that under the 1.5 \textdegree C scenario. This finding explains the feasibility of the relatively weaker climate policies adopted under the 2 \textdegree C scenario, leading to higher gross fossil CO\textsubscript{2} emissions. However, a fair amount of fossil CO\textsubscript{2} emissions is removed in the later period when the costs related to the negative CO\textsubscript{2} emissions are more reasonable. Under the 1.5 \textdegree C scenario, stronger strategies are implemented after the early period. Thus, the gross fossil CO\textsubscript{2} emissions are relatively lower, and the required negative CO\textsubscript{2} emissions do not need to be as high \cite{Fujimori2018}. Here, the general trend can be simply interpreted as that of emit more but reduce more. Moreover, if considering the negative CO\textsubscript{2} emissions, the net forcings are actually lower than the current levels in the energy sector under both the 2 \textdegree C (0.42 Wm\textsuperscript{-2}) and the 1.5 \textdegree C (0.31 Wm\textsuperscript{-2}) scenarios, although gross increases are prominent (\cref*{fig:rfc_grp}a).

Second, in regard to agriculture, the major sources are CH\textsubscript{4} and N\textsubscript{2}O. A considerable amount of the forcings induced by CH\textsubscript{4} and N\textsubscript{2}O still remains under the 2 \textdegree C and 1.5 \textdegree C scenarios (\cref*{fig:rfc_grp}a), due to difficulties in reducing the CH\textsubscript{4} and N\textsubscript{2}O emissions from agriculture \cite{Fujimori2018,Gidden2019} and their relatively long lifetimes, namely, 12.4 years for CH\textsubscript{4} and 121 years for N\textsubscript{2}O (see Table 8.A.1 in ref \cite{ar5ch8}).

Third, the land-use albedo currently exhibits a cooling effect of -$0.16\pm 0.03$ Wm\textsuperscript{-2}. However, the land-use albedo may reveal warming effects in the future, at $0.03\pm 0.08$ Wm\textsuperscript{-2} under the 2 \textdegree C scenario and $0.05\pm 0.13$ Wm\textsuperscript{-2} under the 1.5 \textdegree C scenario (\cref*{fig:rfc_grp}b). The forest cover is expected to increase under the 2 \textdegree C and 1.5 \textdegree C scenarios (\cref*{fig:lc}), which will lower the surface albedo and reflect less of the incoming solar radiation, which in turn will generate lower negative forcings, or more positive forcings, while the current deforestation causes a negative forcing \cite{Myhre2003}. Therefore, to achieve the set climate goals, more forcings need to be reduced from other sources to compensate for this effect. Here, the land-use albedo is estimated by a simple parameterization scheme \cite{Bright2013, Gasser2016} constrained by future land cover changes (see Methods), and the results reveal lower negative forcings, or more positive forcings, than those by the other estimations (\cref*{fig:rfc_LCC}).

{\linespread{1}
\begin{figure}[ht]
  \includegraphics[width=0.8\textwidth]{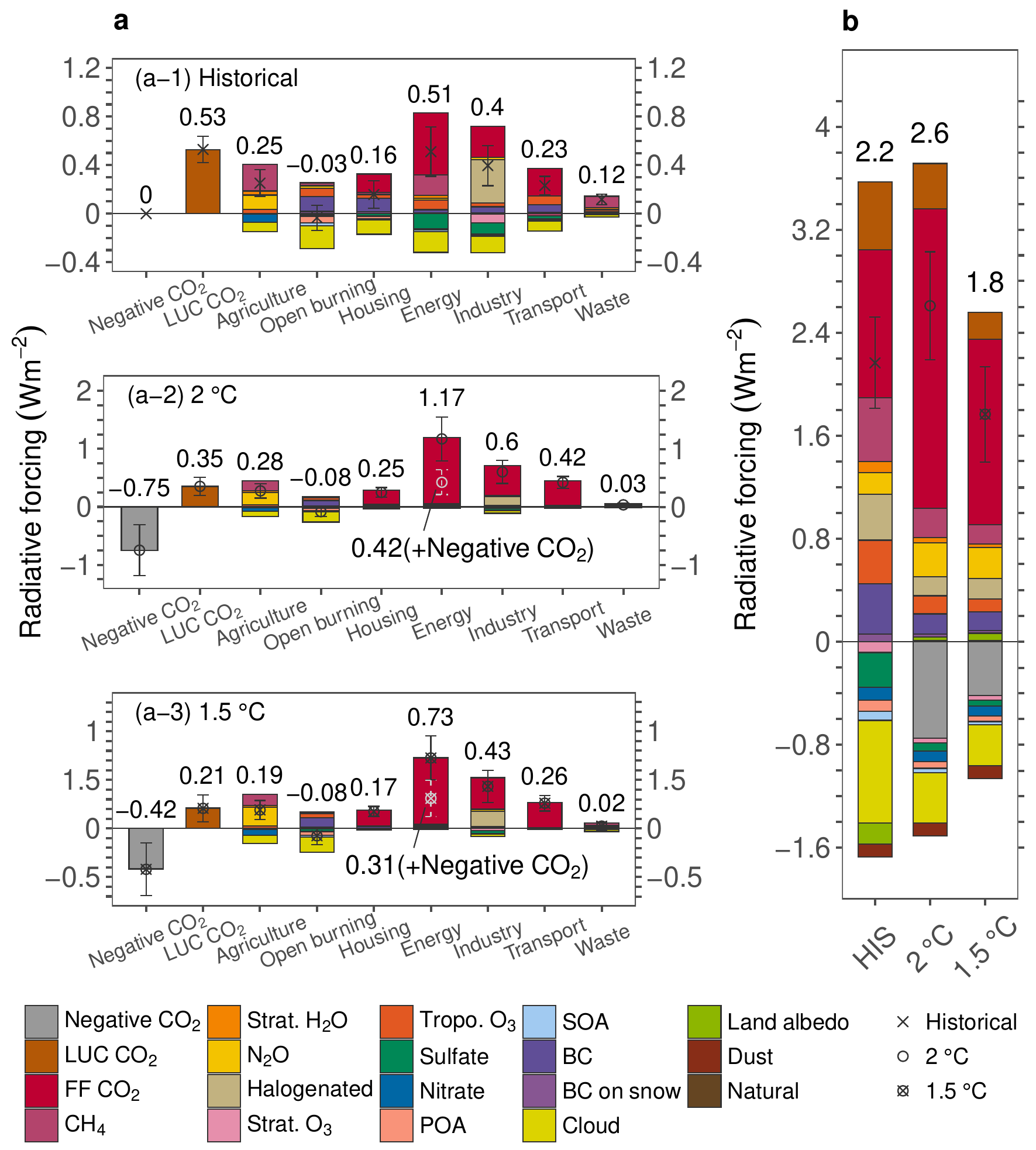}
  \centering
  \caption{\small{\textbf{Contributions of the individual climate forcers.} a, Sectoral contributions decomposed into individual climate forcers from (a-1) the historical period to 2016, (a-2) the 2 \textdegree C climate target by 2100 and (a-3) the 1.5 \textdegree C climate target by 2100. Here, open burning sums the agricultural waste burning, forest burning and grass burning levels. Industry includes industry and solvents, similar to in \cref*{fig:sec_all}. Transport totals the surface transport, aviation and international shipping values. The annotation values under the energy sector in (a-2) and (a-3) denote the forcing values accounting for the negative CO\textsubscript{2} emissions. b, The world total forcings are decomposed into individual climate forcers. The value on top of the bar indicates the mean value of the total radiative forcing, and the error bar indicates the associated uncertainty resulting from the world ensemble. The direct CO\textsubscript{2} emissions are divided into fossil-fuel CO\textsubscript{2} (FF CO\textsubscript{2}), land-use CO\textsubscript{2} (LUC CO\textsubscript{2}), and negative CO\textsubscript{2} emissions, if applicable. The natural forcings include solar irradiance and volcanic activity (\cref*{fig:rfc_natr}). The land-use albedo, mineral dust (Dust) and natural forcings are applied to Fig. b only. The probabilities of reaching the 2 \textdegree C and 1.5 \textdegree C targets here are 56\% and 61\%, respectively (see \cref*{fig:rfc_prob}). All uncertainties are represented as one standard deviation.}}
  \label{fig:rfc_grp}
\end{figure}
}

\subsection{Attributions under the high-emission scenarios}

All potential projected scenarios, including those forcings higher than the 2\textdegree C and 1.5 \textdegree C forcings, are shown in \cref*{fig:rfc_prob}a (for the regional contributions) and \cref*{fig:rfc_prob}b (for the sectoral forcings). We translate the forcing levels into probabilities of exceeding 2\textdegree C or 1.5\textdegree C to demonstrate the likelihood of achieving the climate goals under such conditions. Basically, China, the USA and the EU are still the three major contributors to climate change when high forcings are applied. For example, under the high-forcing scenarios, China may account for approximately 1.1 Wm\textsuperscript{-2}, albeit with greater uncertainty, the USA accounts for approximately 0.9 Wm\textsuperscript{-2}, and the EU accounts for approximately 0.6 Wm\textsuperscript{-2}. All other regions exhibit relatively lower but still significant radiative forcings under the same circumstances, except Japan, where the forcing levels do not greatly change even under the high-forcing scenarios. Sectorally, the energy sector may contribute the highest forcings given its high emissions, up to $3.8$ Wm\textsuperscript{-2}, followed by the industry (up to $1.4$ Wm\textsuperscript{-2}) and transport (up to $1.1$ Wm\textsuperscript{-2}), as well as land-use CO\textsubscript{2} (up to $0.8$ Wm\textsuperscript{-2}). Lower or no negative CO\textsubscript{2} emissions (\cref*{fig:ghg}), relatively fewer nuclear and renewable energy sources (for example, solar and wind), and more fossil-fueled energy use could bring about extremely high climate-related emissions in the energy sector \cite{ONeill2014, Kriegler2017,Fujimori2018, Gidden2019}; hence, high forcings are produced. All sectors reveal high radiative forcing values except for open burning, which remains relatively stable, under the assumed high-forcing scenarios (\cref*{fig:rfc_prob}b).

{\linespread{1}
\begin{figure}[ht]
  \includegraphics[width=\textwidth]{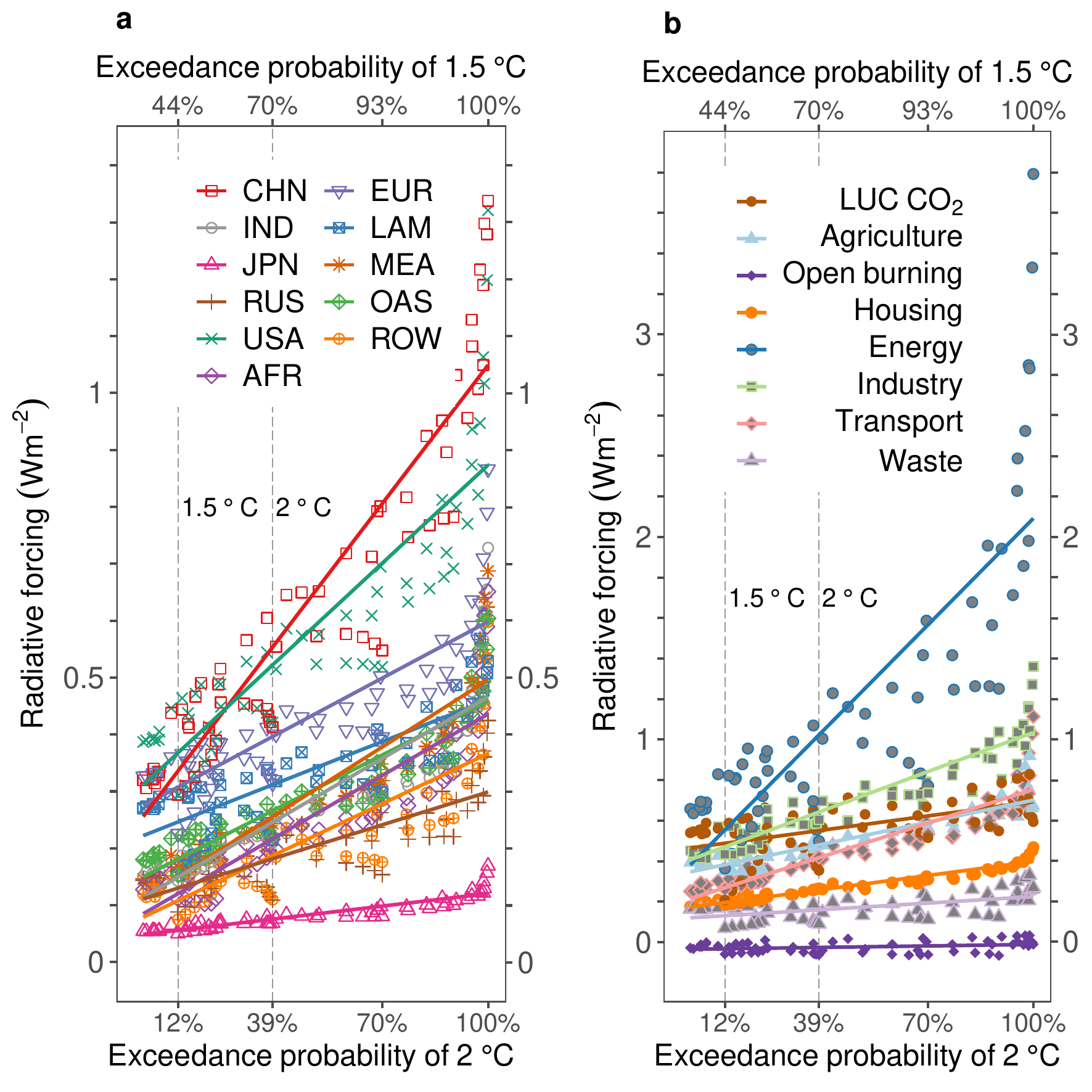}
  \centering
  \caption{\small{\textbf{Regional and sectoral contributions under the different future projections.} a, The relationship between the exceedance probability of 2\textdegree C or 1.5\textdegree C and the regional forcing contributions. The color points are the regional relative contributions. The trends are shown by the colored lines obtained via linear regression. b, The relationship between the exceedance probability of 2\textdegree C or 1.5\textdegree C and the sectoral forcing contributions. The negative CO\textsubscript{2} level is summed into the energy sector to simplify the analysis. The color points are the sectoral forcings. The trends are shown by colored lines obtained through linear regression. The points in Fig. a\&b are sampled (from 2020 to 2100, every 10 years) at seven forcing levels, namely, 1.9 Wm\textsuperscript{-2}, 2.6 Wm\textsuperscript{-2}, 3.4 Wm\textsuperscript{-2}, 4.5 Wm\textsuperscript{-2}, 6.0 Wm\textsuperscript{-2}, 7.0 Wm\textsuperscript{-2} and 8.5 Wm\textsuperscript{-2}. The forcing levels are translated into exceedance probabilities within each sample year. The 1.5\textdegree C and 2\textdegree C results marked with the vertical lines represent the scenarios with forcing levels of 1.9 Wm\textsuperscript{-2} and 2.6 Wm\textsuperscript{-2}, respectively, in 2100.}}
  \label{fig:rfc_prob}
\end{figure}
}

\section{Discussion}

In this study, we applied a simple climate model, SCM4OPT v2.0, to determine the forcing contributions of regions, sectors and climate forcers based on available historical and future projection emissions and land-use datasets. This study provides an IAM-based assessment from a forcing perspective at the sectoral and regional levels. The radiative forcings, including the forcings resulting from the various sources of GHGs, aerosols and pollutants, and land-use albedo, are distinguished among the different regions and sectors. The outputs here can be used to inform policy-makers of the relative importance of the forcing levels resulting from different regional or sectoral sources. 

The results are interpreted with certain caveats and limitations. First, we analyze emission datasets that contain regional and sectoral information, while global-scale datasets are not included. Therefore, our analysis only reflects limited uncertainties that are derived from the available emission estimates. Thus, the analysis here is merely considered as an IAM-based evaluation of the potential future climate, especially under the 2\textdegree C and 1.5 \textdegree C scenarios. Second, the outcome of this study is contingent on the set of selected scenarios used, which are treated equally likely. However, the distribution of total range of future emissions does not necessarily present equal probabilities \cite{Ho2019} (\cref*{fig:gwp_sensi}). We consider a further update for this analysis when the probabilistic emission scenarios are available.

The results showed that the 1.5 \textdegree C target requires most regions and sectors to maintain their forcings not higher than the current levels, while slightly higher future forcing levels than present levels are allowed for the 2 \textdegree C target. The results here can be used to assess the gap between the current and targeted climate levels for both regions and sectors in terms of the radiative forcing. Furthermore, we found that the negative CO\textsubscript{2} forcing is projected to contribute -$0.75\pm 0.44$ Wm\textsuperscript{-2} and -$0.42\pm 0.27$ Wm\textsuperscript{-2} under the 2 \textdegree C and 1.5 \textdegree C scenarios, respectively. Our analysis illustrates the importance of the negative CO\textsubscript{2} emissions in achieving the climate targets from the perspective of the radiative forcing. By using a new land-use forcing parameterization, we further found that less negative forcings, or more positive forcings, for the land-use albedo for the 2 \textdegree C and 1.5 \textdegree C scenarios than those from existing studies. A comprehensive consideration of the available forcing sources is important for the climate change assessment.

\clearpage
\newpage

\section*{Acknowledgments}
This work was supported by the Integrated Research Program for Advancing Climate Models (TOUGOU), Grant Number JPMXD0717935457, from the Ministry of Education, Culture, Sports, Science and Technology (MEXT), Japan. The computing resources were provided by the Japan Agency for Marine-Earth Science and Technology (JAMSTEC). We thank M. Abe for providing the data used for model calibration and T. Gasser for sharing the OSCAR v2.2 source code.

\section*{Author contributions} X.S., K. Tachiiri and K. Tanaka designed the study. X.S. processed the emissions and land cover source data. X.S. developed the model with the help of K. Tanaka and M.W. X.S. performed the calculations and generated the figures. All coauthors contributed to analyzing the results and writing the paper.

\section*{Competing financial interests}
The authors declare that they have no competing financial interests.

\section*{Data availability}
The data used to support the analysis are available from the corresponding author upon reasonable request.

\clearpage
\newpage

\bibliographystyle{iopart-num}
\bibliography{contrib}

\clearpage
\newpage

\setcounter{page}{1}
% \resetlinenumber
% \linenumbers

\begin{center}
  \textbf{\LARGE Supplemental Materials: Source attributions of radiative forcing by regions, sectors, and climate forcers}
  \end{center}
 
  \large{Xuanming Su$^{1\ast}$, Kaoru Tachiiri$^{1,2}$, Katsumasa Tanaka$^{2,3}$, Michio Watanabe$^1$ \& Michio Kawamiya$^{1}$}\\
  \\

\normalsize\noindent{$^{1}$Research Institute for Global Change / Research Center for Environmental Modeling and Application / Earth System Model Development and Application Group, Japan Agency for Marine-Earth Science and Technology (JAMSTEC), Yokohama, Japan}
\normalsize\noindent{$^{2}$Center for Global Environmental Research, National Institute for Environmental Studies (NIES), Tsukuba, Japan}
\normalsize\noindent{$^{3}$Laboratoire des Sciences du Climat et de l’Environnement (LSCE), Commissariat à l’énergie atomique et aux énergies alternatives (CEA), Gif-sur-Yvette, France}

% \normalsize\noindent{$^{4}$Research Institute for Global Change/Research Center for Environmental Modeling and Application/Climate Model Development and Application Group, Japan Agency for Marine-Earth Science and Technology}

% \normalsize\noindent{$^{5}$Research Institute for Global Change/Research Center for Environmental Modeling and Application, Japan Agency for Marine-Earth Science and Technology}

\normalsize\noindent{$^\ast$To whom correspondence should be addressed; e-mail:  suxuanming@jamstec.go.jp.}

\normalsize

\section*{Supplementary materials}

\setcounter{figure}{0}
\makeatletter
\renewcommand{\thefigure}{S\@arabic\c@figure}
\makeatother

\setcounter{table}{0}
\makeatletter
\renewcommand{\thetable}{S\@arabic\c@table}
\makeatother

\setcounter{equation}{0}
\makeatletter
\renewcommand{\thetable}{S\@arabic\c@table}
\makeatother

\begin{figure}[ht]
  \includegraphics[width=0.9\textwidth]{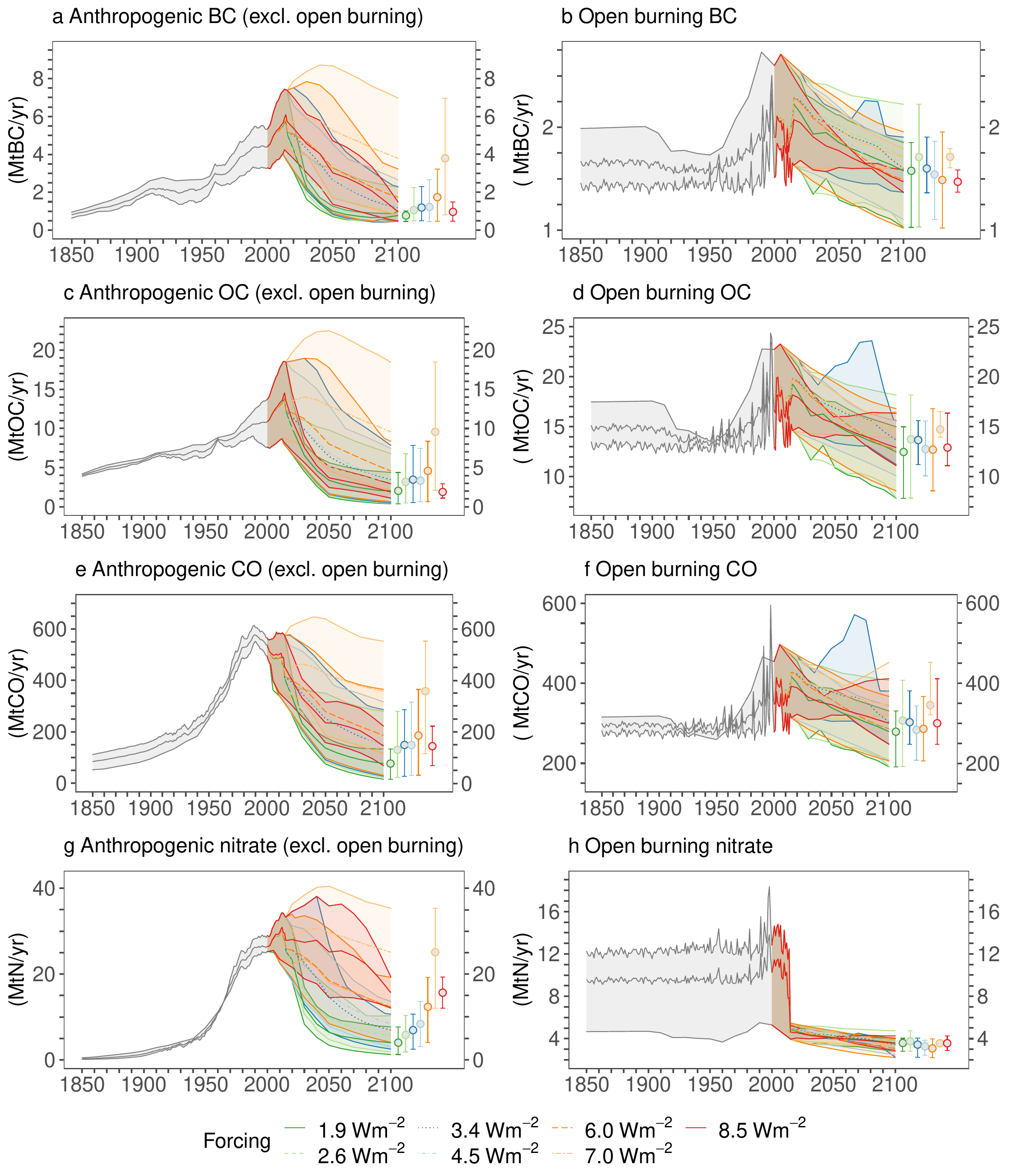}
  \centering
  \caption{\small{\textbf{Historical and future aerosol and pollutant emissions (a-h).} The future projections include seven forcing levels, namely, 1.9 Wm\textsuperscript{-2}, 2.6 Wm\textsuperscript{-2}, 3.4 Wm\textsuperscript{-2}, 4.5 Wm\textsuperscript{-2}, 6.0 Wm\textsuperscript{-2}, 7.0 Wm\textsuperscript{-2} and 8.5 Wm\textsuperscript{-2}. The uncertainty ranges denote the upper and lower trends. The error bars to the right show the upper and lower trends in 2100 at each forcing level. Open burning includes the emissions from agricultural waste burning, forest fires and grassland fires. Sources: the historical emissions stem from ref \cite{Lamarque2009,Gutschow2016,Aardenne2018,Hoesly2018}; future trends come from ref \cite{Fujimori2018,Gidden2019}; open burning originates from ref \cite{Marle2017}.}}
  \label{fig:aero1}
\end{figure}

\begin{figure}[ht]
  \includegraphics[width=0.9\textwidth]{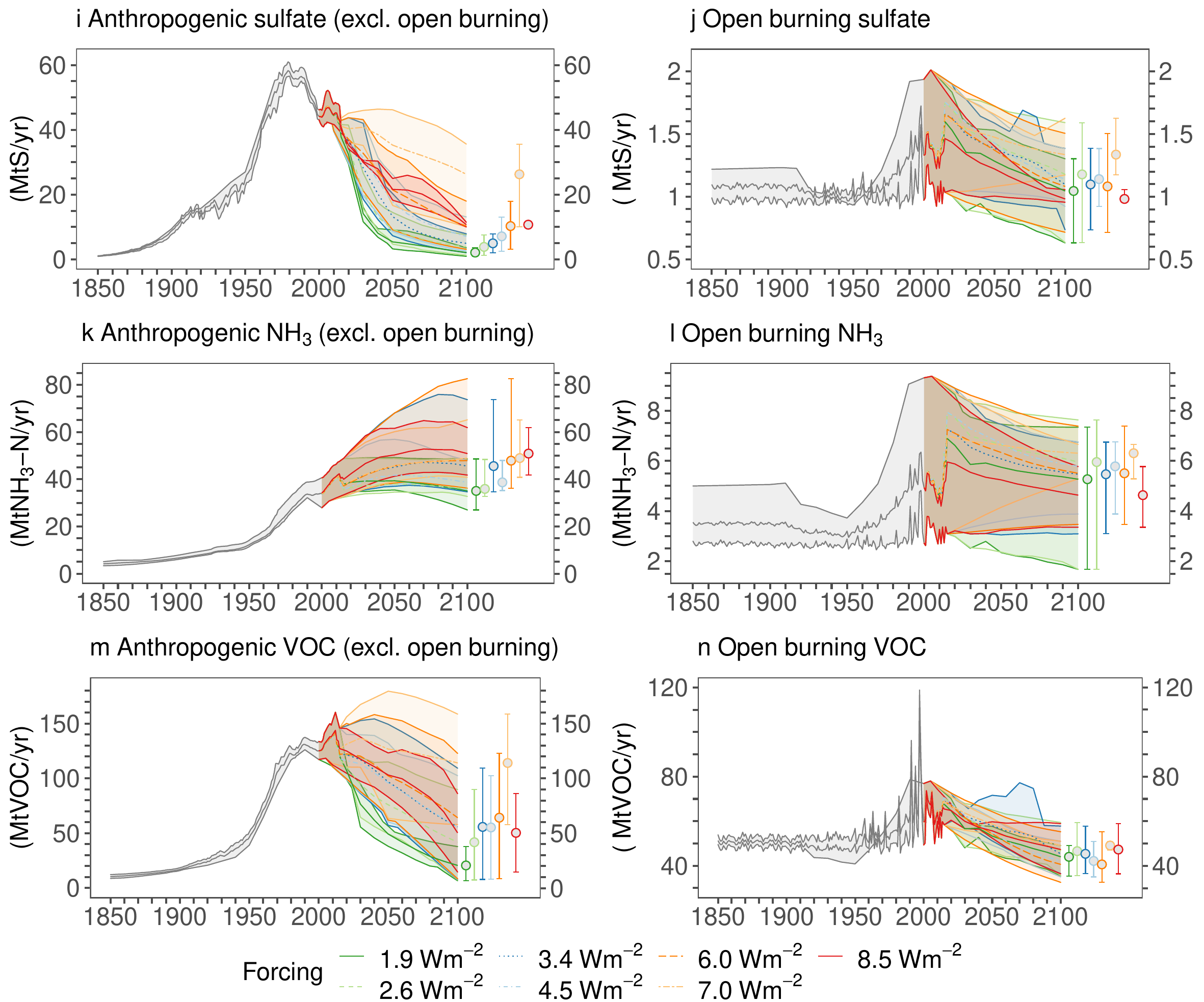}
  \centering
  \caption{\small{\textbf{Historical and future aerosol and pollutant emissions (i-n).} The future projections include seven forcing levels, namely, 1.9 Wm\textsuperscript{-2}, 2.6 Wm\textsuperscript{-2}, 3.4 Wm\textsuperscript{-2}, 4.5 Wm\textsuperscript{-2}, 6.0 Wm\textsuperscript{-2}, 7.0 Wm\textsuperscript{-2} and 8.5 Wm\textsuperscript{-2}. The uncertainty ranges denote the upper and lower trends. The error bars to the right show the upper and lower trends in 2100 at each forcing level. Open burning includes the emissions from agricultural waste burning, forest fires and grassland fires. Sources: the historical emissions are from ref \cite{Lamarque2009,Gutschow2016,Aardenne2018,Hoesly2018}; future trends come from ref \cite{Fujimori2018,Gidden2019}; open burning stems from ref \cite{Marle2017}.}}
  \label{fig:aero2}
\end{figure}

\begin{figure}[ht]
  \includegraphics[width=0.9\textwidth]{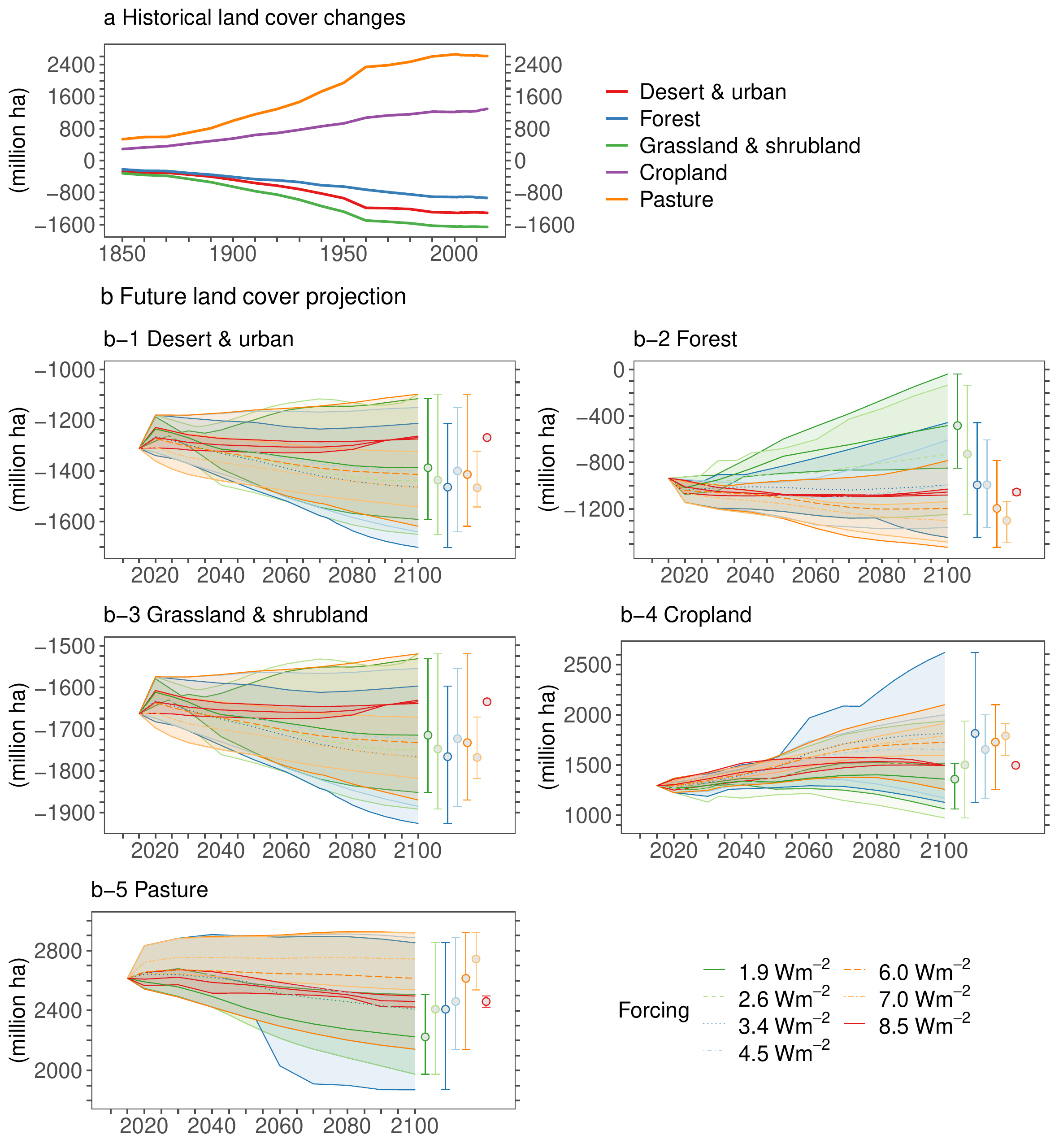}
  \centering
  \caption{\small{\textbf{Historical and future land cover changes, compared to the values in 1700.} The future projections include seven forcing levels, namely, 1.9 Wm\textsuperscript{-2}, 2.6 Wm\textsuperscript{-2}, 3.4 Wm\textsuperscript{-2}, 4.5 Wm\textsuperscript{-2}, 6.0 Wm\textsuperscript{-2}, 7.0 Wm\textsuperscript{-2} and 8.5 Wm\textsuperscript{-2}. The uncertainty ranges denote the upper and lower trends. The error bars to the right show the upper and lower trends in 2100 at each forcing level. Sources: LUH2 v2h\cite{Hurtt2016a}; LUH2 v2f\cite{Hurtt2016b}; AIM-SSP/RCP gridded emission and land-use data\cite{Fujimori2018}.}}
  \label{fig:lc}
\end{figure}

\begin{figure}[ht]
  \centering
  \includegraphics[width=\textwidth]{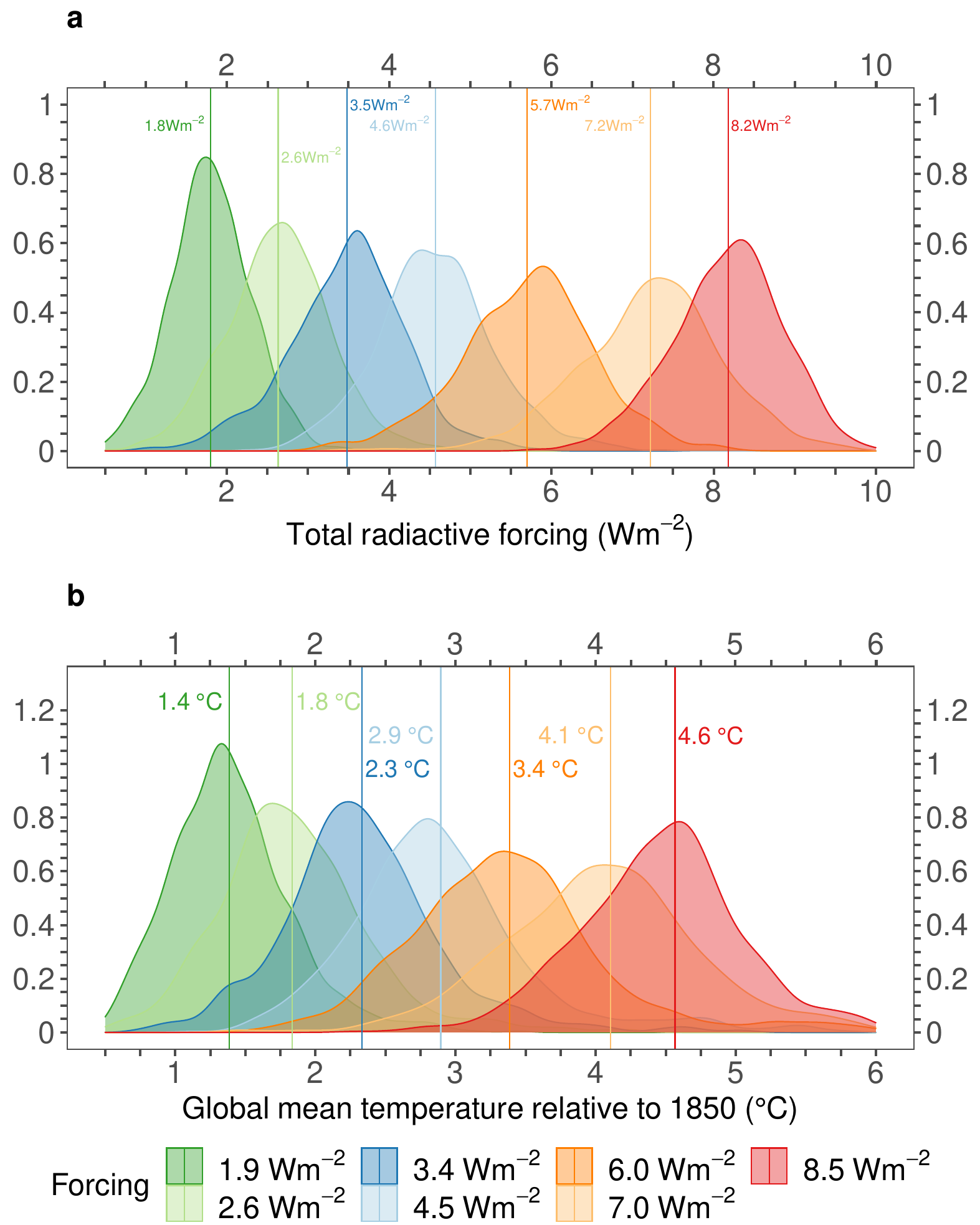}
  \caption{\textbf{Probability distributions of the total radiative forcing and global mean temperature at forcing levels of 1.9 Wm\textsuperscript{-2}, 2.6 Wm\textsuperscript{-2}, 3.4 Wm\textsuperscript{-2}, 4.5 Wm\textsuperscript{-2}, 6.0 Wm\textsuperscript{-2}, 7.0 Wm\textsuperscript{-2} and 8.5 Wm\textsuperscript{-2}, estimated by SCM4OPT v2.0.} a, Total radiative forcing in 2100. b, Global mean temperature increase relative to 1850 in 2100. The color values indicate the mean value at each forcing level.}
  \label{fig:densi}
\end{figure}

\begin{figure}[ht]
  \centering
  \includegraphics[width=\textwidth]{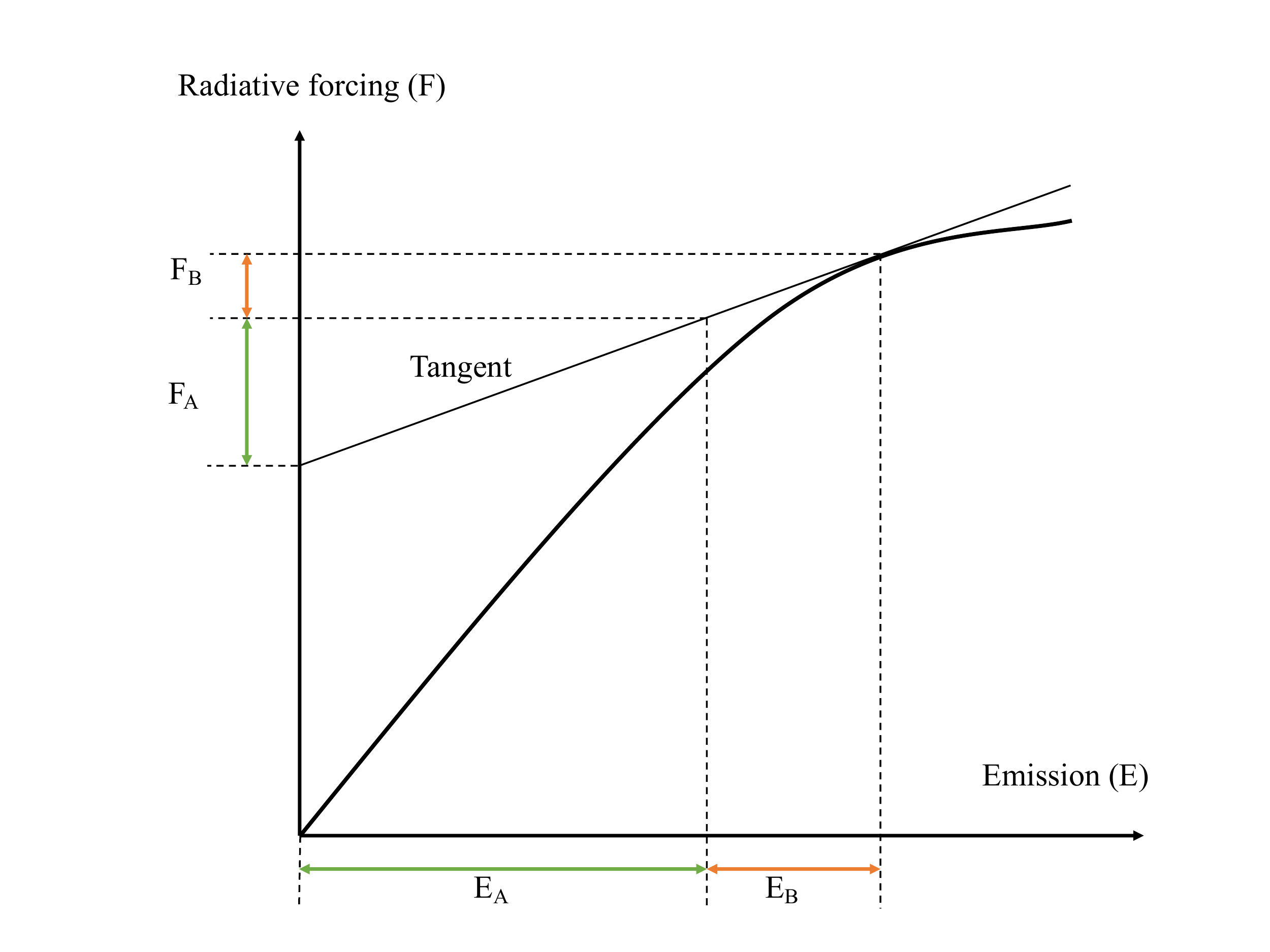}
  \caption{\textbf{The normalized marginal method for the attributions of radiative forcings.} The figure is plotted based on Figure 5 in ref \cite{IPCC2002}.}
  \label{fig:margin}
\end{figure}

\begin{figure}[ht]
  \includegraphics[width=0.9\textwidth]{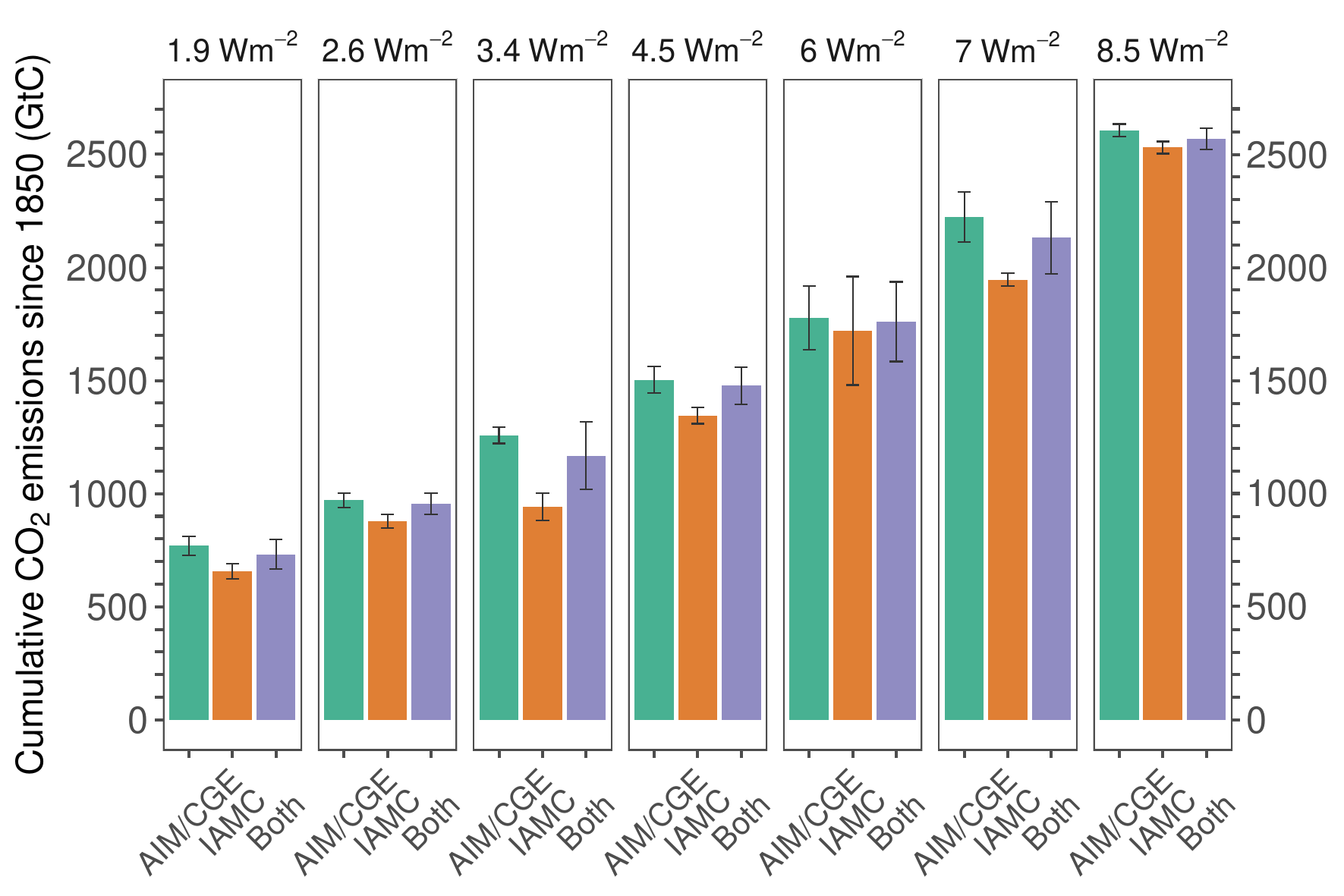}
  \centering
  \caption{\small{\textbf{Cumulative CO\textsubscript{2} emissions projected by AIM/CGE and IAMC.} The uncertainties are represented as one standard deviation. All the AIM/CGE projections are slightly higher than those by the IAMC. This figure shows only the cumulative CO\textsubscript{2} emissions, representing part of the systematic deviations, and variations regarding aerosols and pollutants also occur.}}
  \label{fig:gwp_sensi}
\end{figure}

\begin{itemize}
  \item Model equations for SCM4OPT v2.0
  \item \cref*{tbl_hist} to \cref*{tbl_unc}
  \item \cref*{fig:ghg} to \cref*{fig:gwp_sensi}
\end{itemize}

\clearpage
\newpage

\section*{Model equations for SCM4OPT v2.0}

\subsection*{Carbon cycle}

The perturbation of the atmospheric carbon pool includes the following five components, as defined in \cref*{eq:cflux}.
\begin{enumerate}[label=\protect\circled{\arabic*}]
  \item CO\textsubscript{2} emissions from fossil fuels and industrial sources;
  \item Anthropogenic CO\textsubscript{2} emissions into or removal from the terrestrial biosphere;
  \item CH\textsubscript{4} oxidation of fossil fuels;
  \item Carbon fluxes to or from the terrestrial biosphere due to CO\textsubscript{2} fertilization and climate feedback;
  \item Carbon uptake by oceans.
\end{enumerate}

\begin{equation}
\label{eq:cflux}
\Delta C_{atm}(t)=E^{ind}_{CO_{2}}(t) +E^{lnd}_{CO_{2}}(t)+E_{fCH_{4}}(t)-F_{bio}(t)-F_{ocn}(t)
\end{equation}

\nomenclature[C]{$\Delta C_{atm}(t)$}{Atmospheric carbon pool at time t}
\nomenclature[C]{$E^{ind}_{CO_{2}}(t)$}{CO\textsubscript{2} emissions from fossil fuels and industrial sources}
\nomenclature[C]{$E^{lnd}_{CO_{2}}(t)$}{Anthropogenic CO\textsubscript{2} emissions into or removal from the terrestrial biosphere}
\nomenclature[C]{$E_{fCH_{4}}(t)$}{CH\textsubscript{4} oxidation of fossil fuels}
\nomenclature[C]{$F_{bio}(t)$}{Carbon fluxes to or from the terrestrial biosphere due to CO\textsubscript{2} fertilization and climate feedback}
\nomenclature[C]{$F_{ocn}(t)$}{Carbon uptake by oceans at time t}

\subsubsection*{Terrestrial carbon cycle}

Both the logarithmic and rectangular hyperbolic forms are adopted to simulate the CO\textsubscript{2} fertilization effects. First, the logarithmic description is defined as:
\begin{equation}
\label{eq:ferbeta}
\beta_{log}(t)=1+\beta ln\left(\frac{C_{CO_{2}}(t)}{C_{CO_{2}}^{0}}\right)
\end{equation}

\nomenclature[C]{$\beta$}{CO\textsubscript{2} fertilization factor}
\nomenclature[C]{$C_{CO_{2}}(t)$}{Atmospheric CO\textsubscript{2} concentration at time t}
\nomenclature[C]{$C_{CO_{2}}^{0}$}{Preindustrial CO\textsubscript{2} concentration (278 ppm)}
\nomenclature[C]{$\beta_{log}$}{Fertilization coefficient}

Second, the rectangular hyperbolic description is given by Eq. (\ref{eq:fersigr}-\ref{eq:fersig}):

\begin{equation}
\label{eq:fersigr}
\beta_{sig-r}=\frac{1+\beta \log\left(680 / C_{CO_{2}}^{0}\right)}{1+\beta \log\left(340 / C_{CO_{2}}^{0}\right)}
\end{equation}

\begin{equation}
\label{eq:fersigb}
\beta_{sig-b}=\frac{\left(680-C_{b}\right)-\beta_{sig-r} \left(340-C_{b}\right)}{\left(\beta_{sig-r}-1 \right)\left(680-C_{b}\right) \left(340-C_{b}\right)}
\end{equation}

\begin{equation}
\label{eq:fersig}
\beta_{sig}(t)=\frac{1/\left(C_{CO_{2}}^{0}-C_{b}\right)+\beta_{sig-b}}{1/\left(C_{CO_{2}}(t)-C_{b}\right)+\beta_{sig-b}}
\end{equation}

\nomenclature[C]{$\beta_{sig}(t)$}{Effective CO\textsubscript{2} fertilization factor at time t}
\nomenclature[C]{$C_{b}$}{Concentration at which the net primary productivity (NPP) is zero, which is set to 31 ppm \cite{Gifford1993}}

The CO\textsubscript{2} fertilization coefficient ($\beta_{fert}$) is given by:
\begin{equation}
\label{eq:betafert}
\beta_{fert}(t)=(2-\beta_{m})\beta_{log}(t)+(\beta_{m}-1)\beta_{sig}(t)
\end{equation}

\nomenclature[C]{$\beta_{fert}$}{CO\textsubscript{2} fertilization coefficient}
\nomenclature[C]{$\beta_{m}$}{Allocation coefficient between the two descriptions of the CO\textsubscript{2} fertilization effects}

The NPP ($F_{NPP}(t)$) and heterotrophic respiration ($F_{rsp}(t)$) are defined as products of the initial carbon flux and a certain fertilization coefficient, considering an exponential temperature feedback effect (Eq. \ref{eq:fnpp} and \ref{eq:fresp}).

\begin{equation}
\label{eq:fnpp}
F_{NPP}(t) = F^{0}_{NPP}   \beta_{fert}(t) \exp\left(\sigma_{NPP} \Delta T(t)\right)
\end{equation}

\begin{equation}
\label{eq:fresp}
F_{rsp}(t)=F^{0}_{rsp}  \beta_{fert}(t) \exp\left(\sigma_{rsp} \Delta T(t)\right)
\end{equation}

\nomenclature[C]{$F_{NPP}(t)$}{Net primary productivity (NPP) at time t}
\nomenclature[C]{$F_{rsp}(t)$}{Heterotrophic respiration at time t}
\nomenclature[C]{$F^{0}_{rsp}$}{Preindustrial heterotrophic respiration}
\nomenclature[C]{$\sigma_{rsp}$}{Sensitivity to changes in temperature}

The gross land-use emission levels are defined as the sums of the net land-use emissions and the corresponding regrowth, as shown in Eq. (\ref{eq:dpgross}-\ref{eq:dsgross}):

\begin{equation}
\label{eq:dpgross}
D^{gross}_{P}(t)=E^{lnd}_{P}(t) + G_{P}(t)
\end{equation}

\begin{equation}
\label{eq:dhgross}
D^{gross}_{H}(t)=E^{lnd}_{H}(t) + G_{H}(t)
\end{equation}

\begin{equation}
\label{eq:dsgross}
D^{gross}_{S}(t)=E^{lnd}_{S}(t) + G_{S}(t)
\end{equation}

\nomenclature[C]{$D^{gross}_{i}(t)$}{Gross land-use emission level, $i \in \{P, H, S\} $ denote the living plant pool, the detritus pool and the soil pool, respectively}
\nomenclature[C]{$E^{lnd}_{i}(t)$}{Net land-use emission level, $i \in \{P, H, S\} $ denote the living plant pool, the detritus pool and the soil pool, respectively}
\nomenclature[C]{$G_{i}(t)$}{Carbon flux originating from regrowth, $i \in \{P, H, S\} $ denote the living plant pool, the detritus pool and the soil pool, respectively}

Proportions of the net land-use emission levels are allocated as:

\begin{enumerate}[label=\protect\circled{\arabic*}]
  \item Living plant pool;
  \item Detritus pool;
  \item Soil pool.
\end{enumerate}

Please refer to Eq. (\ref{eq:elndp}-\ref{eq:elnds}).

\begin{equation}
\label{eq:elndp}
E^{lnd}_{P}(t)=\delta_{P}   E^{lnd}_{CO_{2}}(t)
\end{equation}

\begin{equation}
\label{eq:elndh}
E^{lnd}_{H}(t)=\delta_{H}   E^{lnd}_{CO_{2}}(t)
\end{equation}

\begin{equation}
\label{eq:elnds}
E^{lnd}_{S}(t)=\delta_{S}   E^{lnd}_{CO_{2}}(t)
\end{equation}

% \nomenclature[C]{$E^{lnd}_{CO_{2}}(t)$}{Net land-use emission level}
\nomenclature[C]{$E^{lnd}_{P}(t)$}{Living plant pool}
\nomenclature[C]{$E^{lnd}_{H}(t)$}{Detritus pool}
\nomenclature[C]{$E^{lnd}_{S}(t)$}{Soil pool}
\nomenclature[C]{$\delta_{i}$}{Land-use emission distribution factors}

The regrowth here is defined to be linearly related to the relaxation time.

\begin{equation}
\label{eq:gp}
G_{P}(t)=a_{P}+b_{P} \tau_{P}(t)
\end{equation}

\begin{equation}
\label{eq:gh}
G_{H}(t)=a_{H}+b_{H} \tau_{H}(t)
\end{equation}

\begin{equation}
\label{eq:gs}
G_{S}(t)=a_{S}+b_{S} \tau_{S}(t)
\end{equation}

\nomenclature[C]{$G_{i}(t)$}{Land-use regrowth, $a_{i}$ and $b_{i}$ are parameters that are estimated based on CMIP5 outputs}
\nomenclature[C]{$\tau_{i}(t)$}{Regrowth relaxation time, $a_{i}$ and $b_{i}$ are parameters that are estimated based on CMIP5 outputs}

The relaxation times are defined as follows:

\begin{equation}
\label{eq:taup}
\tau_{P}(t)=\frac{P_{0}-\psi \int_{0}^{t} E^{lnd}_{P}(t')dt'}{dP_{0}}
\end{equation}

\begin{equation}
\label{eq:tauh}
\tau_{H}(t)=\frac{H_{0}-\psi \int_{0}^{t} E^{lnd}_{H}(t')dt'}{dH_{0}}
\end{equation}

\begin{equation}
\label{eq:taus}
\tau_{S}(t)=\frac{S_{0}-\psi \int_{0}^{t} E^{lnd}_{S}(t')dt'}{dS_{0}}
\end{equation}

\nomenclature[C]{$P_{0}$, $H_{0}$ and $S_{0}$}{Initial states of the living plant pool, the detritus pool and the soil pool, respectively}
\nomenclature[C]{$\psi$}{Fraction of the gross deforestation without regrowth}
\nomenclature[C]{$dP_{0}$, $dH_{0}$ and $dS_{0}$}{Initial decay rates}

Therefore, the annual decay rates for the living plant pool, detritus pool and soil pool are defined as shown in Eq. (\ref{eq:dp}-\ref{eq:ds}):
\begin{equation}
\label{eq:dp}
dP(t)=C_{P}(t)   \frac{1}{\tau_{P}(t)}
\end{equation}

\begin{equation}
\label{eq:dH}
dH(t)=C_{H}(t)   \frac{1}{\tau_{H}(t)}   \exp\left(\sigma_{H} \Delta T(t)\right)
\end{equation}

\begin{equation}
\label{eq:ds}
dS(t)=C_{S}(t) \frac{1}{\tau_{S}(t)} \exp\left(\sigma_{S} \Delta T(t)\right)
\end{equation}

\nomenclature[C]{$C_{P}(t)$, $C_{H}(t)$ and $C_{S}(t)$}{Amounts of carbon remaining in the living plant pool, detritus pool and soil pool, respectively}
\nomenclature[C]{$\sigma_{H}$ and $\sigma_{S}$}{Temperature feedback coefficients for the detritus pool and soil pool, respectively}

The perturbations of carbon in the living plant pool, detritus pool and soil pool at time t are defined as shown in Eq. (\ref{eq:deltacp}-\ref{eq:deltacs}):

\begin{equation}
\label{eq:deltacp}
\Delta P(t)=F_{NPP}(t)   \nu_{P} - dP(t) - D^{gross}_{P}(t) - F_{rsp}(t)
\end{equation}

\begin{equation}
\label{eq:deltacd}
\Delta H(t)=F_{NPP}(t)   \nu_{H} - dH(t) - D^{gross}_{H}(t) + dP(t)   \rho_{p2d}
\end{equation}

\begin{equation}
\label{eq:deltacs}
\Delta S(t)=F_{NPP}(t)   \nu_{S} - dS(t) - D^{gross}_{S}(t) + dP(t)   \rho_{p2s} + dH(t)   \delta_{d2s}
\end{equation}

\nomenclature[C]{$\Delta P(t)$, $\Delta H(t)$ and $\Delta S(t)$}{Total changes in the carbon levels for the living plant pool, the detritus pool and the soil pool, respectively}
\nomenclature[C]{$\nu_{P}$, $\nu_{H}$ and $\nu_{S}$=1-$\nu_{P}$-$\nu_{H}$}{NPP partition factors for the living plant pool, the detritus pool and the soil pool, respectively}
\nomenclature[C]{$\rho_{p2d}$ and $\rho_{p2s}$=1-$\rho_{p2d}$}{Fractions of $dP(t)$ that are distributed to the detritus and soil pools, respectively}
\nomenclature[C]{$\delta_{d2s}$}{Fraction of $dH(t)$ going to the soil pool}

Therefore, the carbon flux to or from the terrestrial biosphere can be calculated as follows:

\begin{equation}
\label{eq:fbio}
F_{bio}(t)=\Delta P(t)+\Delta H(t)+\Delta S(t)
\end{equation}

\nomenclature[C]{$F_{bio}(t)$}{Carbon flux to or from the terrestrial biosphere}

Here, we fitted the land net primary productivity (NPP), land surface net downward carbon flux (NBP), ocean surface downward carbon flux (fgco2) and CO\textsubscript{2} concentration of SCM4OPT v2.0 to the outputs of three CMIP5 experiments, namely, the historical, RCP26 and RCP85 experiments. The calibration procedures were performed in several steps, thereby minimizing the sum of squared errors (SSEs) with the associated variables.

\subsubsection*{Oceanic carbon cycle}

We apply the method proposed by \cite{Hartin2015, Hartin2016} to construct the oceanic carbon cycle (Eq.\ref{eq:dic}-\ref{eq:focn}).

\begin{align}
  \label{eq:dic}
  &DIC(obx,t)\cdot\left(\frac{K_{1}(obx,t)}{H(obx,t)} + 2\frac{K_{1}(obx,t)K_{2}(obx,t)}{H(obx,t)^{2}}\right) = \nonumber\\
   &\left(ALK(obx,t) - \frac{K_{B}(obx,t)BOR(obx)}{K_{B}(obx,t) + H(obx,t)} - \frac{K_{W}(obx,t)}{H(obx,t)} + H(obx,t) \right)\cdot \nonumber \\
   &\left(1 + \frac{K_{1}(obx,t)}{H(obx,t)} + \frac{K_{1}(obx,t)K_{2}(obx,t)}{H(obx,t)^{2}} \right)
\end{align}

\nomenclature[C]{$DIC(obx,t)$}{Dissolved inorganic fraction for ocean box $obx$ and time $t$}
\nomenclature[C]{$K_{1}(obx,t)$}{First acidity constant of carbonic acid for ocean box $obx$ and time $t$}
\nomenclature[C]{$K_{2}(obx,t)$}{Second acidity constant of carbonic acid for ocean box $obx$ and time $t$}
\nomenclature[C]{$H(obx,t)$}{Concentration of [H\textsuperscript{+}] for ocean box $obx$ and time $t$}
\nomenclature[C]{$ALK(obx,t)$}{Total alkalinity for ocean box $obx$ and time $t$}
\nomenclature[C]{$K_{B}(obx,t)$}{Dissociation constant of boric acid for ocean box $obx$ and time $t$}
\nomenclature[C]{$BOR(obx)$}{Total boron level for ocean box $obx$}
\nomenclature[C]{$K_{W}(obx,t)$}{Dissociation constant of water for ocean box $obx$ and time $t$}

\begin{equation}
  \label{eq:co2sys}
  CO^{sys}_{2}(obx,t) = \frac{DIC(obx,t)}{1 + \frac{K_{1}(obx,t)}{H(obx,t)} + \frac{K_{1}(obx,t)K_{2}(obx,t)}{H(obx,t)^{2}}}
\end{equation}

\nomenclature[C]{$CO^{sys}_{2}(obx,t)$}{Dissolved inorganic (DIC) fraction of the system for ocean box $obx$ and time $t$}

\begin{equation}
  \label{eq:pco2}
  pCO_{2}(obx,t) = \frac{CO^{sys}_{2}(obx,t)}{K_{H}(obx,t)}
\end{equation}

\nomenclature[C]{$pCO_{2}(obx,t)$}{Sea surface partial pressure for ocean box $obx$ and time $t$}
\nomenclature[C]{$K_{H}(obx,t)$}{Henry’s constant for ocean box $obx$ and time $t$}

\begin{equation}
  \label{eq:hco3}
  HCO_{3}(obx,t) = \frac{DIC(obx,t)}{1 + \frac{H(obx,t)}{K_{1}(obx,t)} + \frac{K_{2}(obx,t)}{H(obx,t)}}
\end{equation}

\nomenclature[C]{$HCO_{3}(obx,t)$}{Concentration of ocean bicarbonate $HCO_{3}^{-}$ for ocean box $obx$ and time $t$}

\begin{equation}
  \label{eq:co3}
  CO_{3}(obx,t) = \frac{DIC(obx,t)}{1+\frac{H(obx,t)}{K_{2}(obx,t)} + \frac{H(obx,t)^{2}}{K_{1}(obx,t)K_{2}(obx,t)}}
\end{equation}

\nomenclature[C]{$CO_{3}(obx,t)$}{Concentration of ocean carbonate $CO_{3}^{2-}$ for ocean box $obx$ and time $t$}

\begin{equation}
  \label{eq:k1}
  K_{1}(obx,t) = \frac{H(obx,t)HCO_{3}(obx,t)}{CO^{sys}_{2}(obx,t)}
\end{equation}

\begin{equation}
  \label{eq:k2}
  K_{2}(obx,t) = \frac{H(obx,t)CO_{3}(obx,t)}{HCO_{3}(obx,t)}
\end{equation}

\begin{equation}
  \label{eq:kb}
  K_{B}(obx,t) = \frac{H(obx,t)BOH_{4}(obx)}{BOH_{3}(obx)}
\end{equation}

\begin{equation}
  \label{eq:bor}
  BOR(obx) = 416.0 \cdot \frac{S}{35.0} = BOH_{4}(obx) + BOH_{3}(obx)
\end{equation}

\nomenclature[C]{$BOH_{4}(obx)$}{Ocean borate level}
\nomenclature[C]{$BOH_{3}(obx)$}{Ocean boric acid level}

\begin{equation}
  \label{eq:kw}
  K_{W}(obx,t) = \frac{H(obx,t)}{OH(obx,t)}
\end{equation}

\nomenclature[C]{$OH(obx,t)$}{Concentration of $OH^{-1}$}

\begin{equation}
  \label{eq:fas}
  F_{as}(obx,t) = \kappa_{s}\alpha_{s}\cdot\left(C_{CO_{2}}(t)-pCO_{2}(obx,t)\right)
\end{equation}

\nomenclature[C]{$F_{as}(obx,t)$}{Carbon fluxes between the atmosphere and surface ocean box for ocean box $obx$ and time $t$, if applicable}
\nomenclature[C]{$\kappa_{s}$}{CO\textsubscript{2} transfer velocity}
\nomenclature[C]{$\alpha_{s}$}{Solubility of CO\textsubscript{2} in seawater}

\begin{equation}
  \label{eq:focn}
  F_{ocn}(t) = \sum_{obx} F_{as}(obx,t)
\end{equation}

The carbon in the atmospheric pool is converted into the CO\textsubscript{2} concentration by:

\begin{equation}
\label{eq:cco2}
C_{CO_{2}}(t)=\frac{C_{atm}(t)}{\alpha_{ppm2gtc}}
\end{equation}

\nomenclature[C]{$C_{CO_{2}}(t)$}{CO\textsubscript{2} concentration}
\nomenclature[C]{$\alpha_{ppm2gtc}$}{Unit conversion factor from ppm into Gt C, = 2.123 Gt C ppm\textsuperscript{-1}}

The radiative forcing from CO\textsubscript{2} can be obtained as:

\begin{equation}
\label{eq:fco2}
f_{CO_{2}}(t)=\alpha_{CO_{2}} \log\frac{C_{CO_{2}}(t)}{C^{0}_{CO_{2}}}
\end{equation}
\nomenclature[C]{$f_{CO_{2}}(t)$}{CO\textsubscript{2} radiative forcing at time $t$}
\nomenclature[C]{$\alpha_{CO_{2}}$}{Forcing scaling parameter, = $\frac{3.71}{\log\left(2\right)}$=5.35 Wm\textsuperscript{-2} \cite{Myhre1998}}
% \nomenclature[C]{$C^{0}_{CO_{2}}$}{Initial CO\textsubscript{2} concentration}

\subsection*{CH\textsubscript{4}}
The change in the CH\textsubscript{4} concentration is directly calculated from the CH\textsubscript{4} emissions from natural, industrial and land-use sources and from the CH\textsubscript{4} sinks in the troposphere (based on the lifetime of OH), stratosphere, and soil.

\begin{equation}
\label{eq:deltach4}
\Delta C_{CH_{4}}(t) = \frac{E^{nat}_{CH_{4}} + E^{ind}_{CH_{4}}(t) + E^{lnd}_{CH_{4}}(t)}{\theta_{CH_{4}}}- \frac{C_{CH_{4}}(t-1)}{\tau^{tot}_{CH_{4}}(t-1)}
\end{equation}

\nomenclature[G]{$\Delta C_{CH_{4}}(t)$}{Change in the CH\textsubscript{4} concentration at time $t$}
\nomenclature[G]{$E^{nat}_{CH_{4}}$}{Natural CH\textsubscript{4} emissions, =274.5 Mt CH\textsubscript{4} yr\textsuperscript{-1}}
\nomenclature[G]{$E^{ind}_{CH_{4}}(t)$}{Industrial CH\textsubscript{4} emissions at time $t$}
\nomenclature[G]{$E^{lnd}_{CH_{4}}(t)$}{Land-use source CH\textsubscript{4} emissions at time $t$}
\nomenclature[G]{$\theta_{CH_{4}}$}{CH\textsubscript{4} conversion factor, 2.78 Tg ppb\textsuperscript{-1}}

\nomenclature[G]{$\tau^{tot}_{CH_{4}}(t)$}{CH\textsubscript{4} lifetime at time $t$}

\begin{equation}
\label{eq:tautot}
\frac{1}{\tau^{tot}_{CH_{4}}(t)} = \frac{1}{\tau^{init}_{CH_{4}}/\tau^{rel}_{OH}(t)} + \frac{1}{\tau^{soil}_{CH_{4}}} + \frac{1}{\tau^{oth}_{CH_{4}}}
\end{equation}

\nomenclature[G]{$\tau^{init}_{CH_{4}}$}{Initial lifetime of OH, =9.6 years}
\nomenclature[G]{$\tau^{init}_{CH_{4}}/\tau^{rel}_{OH}(t)$}{CH\textsubscript{4} lifetime in the troposphere}
\nomenclature[G]{$\tau^{soil}_{CH_{4}}$}{CH\textsubscript{4} lifetime in soil, =160 years}
\nomenclature[G]{$\tau^{oth}_{CH_{4}}$}{CH\textsubscript{4} lifetime in the stratosphere, =120 years}

The change in the tropospheric OH abundance relative to the level in 2000 is thus modeled as:

\begin{align}
\label{eq:reloh}
\tau^{rel}_{OH}(t)=&\quad S_{\tau_{CH_{4}}} \Delta T_{2k}(t) +\left(\frac{C_{CH_{4}}(t)}{C^{2k}_{CH_{4}}}\right)^{S^{OH}_{CH_{4}}} \nonumber\\
                &\cdot\exp\left(S^{OH}_{NO_{x}} \Delta E_{NO_{x}}(t)+S^{OH}_{CO} \Delta E_{CO}(t)+S^{OH}_{VOC} \Delta E_{VOC}(t)\right)
\end{align}

\nomenclature[G]{$S^{OH}_{x}$}{Sensitivities of the tropospheric OH to CH\textsubscript{4}, NO\textsubscript{x}, CO and VOC, with values of -0.32, +0.0042, -1.05E-4 and -3.15E-4, respectively}
\nomenclature[G]{$C^{2k}_{CH_{4}}$}{CH\textsubscript{4} concentration in 2000}
\nomenclature[G]{$S_{\tau_{CH_{4}}}$}{CH\textsubscript{4} temperature sensitivity coefficient of the tropospheric chemical reactions, =0.0316 \degree C\textsuperscript{-1} \cite{Meinshausen2011}}
\nomenclature[G]{$\Delta T_{2k}(t)$}{Temperature change above the 2000 level}

\subsection*{N\textsubscript{2}O}

The feedback effect of the atmospheric N\textsubscript{2}O concentration on its own lifetime is approximated as:

\begin{equation}
\label{eq:taun2o}
\tau_{N_{2}O}(t) = \tau^{init}_{N_{2}O} \left(\frac{C_{N_{2}O}(t)}{C^{2k}_{N_{2}O}}\right)^{S_{\tau_{N_{2}O}}}
\end{equation}

\nomenclature[G]{$C_{N_{2}O}(t)$}{N\textsubscript{2}O concentration}
\nomenclature[G]{$\tau_{N_{2}O}(t)$}{N\textsubscript{2}O lifetime at time $t$}
\nomenclature[G]{$\tau^{init}_{N_{2}O}$}{Initial N\textsubscript{2}O lifetime, =120 years}
\nomenclature[G]{$C^{2k}_{N_{2}O}$}{N\textsubscript{2}O concentration in 2000}
\nomenclature[G]{$S_{\tau_{N_{2}O}}$}{N\textsubscript{2}O sensitivity coefficient, =-0.05}

The change in the atmospheric N\textsubscript{2}O concentration is calculated as:

\begin{equation}
\label{eq:deltan2o}
\Delta C_{N_{2}O}(t) = \frac{E^{nat}_{N_{2}O} + E^{ind}_{N_{2}O}(t) + E^{lnd}_{N_{2}O}(t)}{\theta_{N_{2}O}}- \frac{C_{N_{2}O}(t-1)}{\tau_{N_{2}O}(t-1)}
\end{equation}

\nomenclature[G]{$\Delta C_{N_{2}O}(t)$}{N\textsubscript{2}O concentration change at time $t$}
\nomenclature[G]{$E^{nat}_{N_{2}O}$}{Natural N\textsubscript{2}O emissions, =8.4 Mt N\textsubscript{2}O-N yr\textsuperscript{-1}}
\nomenclature[G]{$E^{ind}_{N_{2}O}(t)$}{Industrial N\textsubscript{2}O emissions at time $t$}
\nomenclature[G]{$E^{lnd}_{N_{2}O}(t)$}{Land-use source N\textsubscript{2}O emissions at time $t$}
\nomenclature[G]{$\theta_{N_{2}O}$}{N\textsubscript{2}O conversion factor, =4.81 Tg ppb\textsuperscript{-1}}

The radiative forcings of CH\textsubscript{4} ($f_{CH_{4}}(t)$) and N\textsubscript{2}O ($f_{N_{2}O}(t)$) are calculated following the standard IPCC (2001) methods \cite{ipcc2001ch6}, as shown in Eq. (\ref{eq:fch4}-\ref{eq:fmn}):

\begin{align}
\label{eq:fch4}
f_{CH_{4}}(t)=&\alpha_{CH_{4}}\left(\sqrt{C_{CH_{4}}(t)}-\sqrt{C^{0}_{CH_{4}}}\right) - \nonumber\\
&\left(f_{mn}\left(C_{CH_{4}}(t),C^{0}_{N_{2}O}\right)-f_{mn}\left(C^{0}_{CH_{4}},C^{0}_{N_{2}O}\right)\right)
\end{align}

\begin{align}
\label{eq:fn2o}
f_{N_{2}O}(t)=&\alpha_{N_{2}O}\left(\sqrt{C_{N_{2}O}(t)}-\sqrt{C^{0}_{N_{2}O}}\right)- \nonumber\\
&\left(f_{mn}\left(C^{0}_{CH_{4}},C_{N_{2}O}(t)\right)-f_{mn}\left(C^{0}_{CH_{4}},C^{0}_{N_{2}O}\right)\right)
\end{align}

\nomenclature[G]{$\alpha_{CH_{4}}$}{CH\textsubscript{4} scaling factor, =0.036}
\nomenclature[G]{$\alpha_{N_{2}O}$}{N\textsubscript{2}O scaling factor, =0.12}
\nomenclature[G]{$C^{0}_{CH_{4}}$}{CH\textsubscript{4} preindustrial concentration, =721.9 ppb}
\nomenclature[G]{$C^{0}_{N_{2}O}$}{N\textsubscript{2}O preindustrial concentration, =273.0 ppb}

The function $f_{mn}(M,N)$ defining the overlap between CH\textsubscript{4} and N\textsubscript{2}O is:
\begin{align}
\label{eq:fmn}
f_{mn}(M,N) =&  \nonumber \\
& 0.47 \log\left(1+0.6356\left(\frac{MN}{10^{6}}\right)^{0.75}+0.007 \frac{M}{10^{3}} \left(\frac{MN}{10^{6}}\right)^{1.52}\right)
\end{align}

\nomenclature[G]{$M$ and $N$}{CH\textsubscript{4} and N\textsubscript{2}O concentration inputs}

\subsection*{Halogenated gases}

All the available halogenated gases are treated separately with regard to their concentrations \cite{Meinshausen2011, Hartin2015}:

\begin{align}
\label{eq:chc}
C_{hc}(t+1,hc) =&\tau_{hc}\frac{E(t,hc)}{\mu_{hc}} \frac{\rho_{atm}}{m_{atm}}\cdot \nonumber \\
 & \left(1-\exp(-\frac{1}{\tau_{hc}})\right)+C_{hc}(t,hc)\left(1-\exp(-\frac{1}{\tau_{hc}})\right)
\end{align}

\nomenclature[G]{$C_{hc}(t+1,hc)$}{Concentration (in ppt) of halogenated gas $hc$ in year $t+1$}
\nomenclature[G]{$E(t,hc)$}{Halogenated gas emission level of $hc$ in kt yr\textsuperscript{-1}}
\nomenclature[G]{$\mu_{hc}$}{Molar mass of halogenated gas $hc$}
\nomenclature[G]{$\tau_{hc}$}{Lifetime of halogenated gas $hc$}
\nomenclature[G]{$\rho_{atm}$}{Average density of air}
\nomenclature[G]{$m_{atm}$}{Total mass of the atmosphere}

The radiative forcing from each halogenated gas is given by:

\begin{equation}
\label{eq:fhc}
f_{hc}(t,hc)=\alpha_{hc}\left(C_{hc}(t,hc)-C^{0}_{hc}\right)
\end{equation}

\nomenclature[G]{$f_{hc}(t,hc)$}{Halogenated gas radiative forcing}
\nomenclature[G]{$\alpha_{hc}$}{Halogenated gas radiative efficiency}
\nomenclature[G]{$C^{0}_{hc}$}{Halogenated gas preindustrial atmospheric concentration}

\subsection*{Direct effect of aerosols}
We update the estimation of the direct effects from aerosols based on \cite{Gasser2016}. The change in the sulfate burden is assessed to capture the radiative forcing impacts resulting from sulfate aerosols.

\begin{align}
\label{eq:so4}
C_{SO_{4}}(t) = &C_{SO_{4}}^{0} + \alpha_{SO_{4}} \tau_{SO_{2}} \left(E_{SO_{2}}^{ind}(t) + E_{SO_{2}}^{lnd}(t) \right) + \nonumber \\
&\alpha_{SO_{4}} \tau_{dms} E_{dms}(t) + \Gamma_{SO_{4}} \Delta T_{as}(t)
\end{align}

\nomenclature[A]{$C_{SO_{4}}(t)$}{Sulfate concentration at time $t$}
\nomenclature[A]{$C_{SO_{4}}^{0}$}{Initial sulfate concentration}
\nomenclature[A]{$\alpha_{SO_{4}}$}{Conversion of $SO_{4}$ from Tg S into Tg (SO4)}
\nomenclature[A]{$\tau_{SO_{2}}$}{Lifetime of $SO_{2}$}
\nomenclature[A]{$E_{SO_{2}}^{ind}(t)$}{Industrial $SO_{2}$ emissions at time $t$}
\nomenclature[A]{$E_{SO_{2}}^{lnd}(t)$}{Land-use $SO_{2}$ emissions at time $t$}
\nomenclature[A]{$\tau_{dms}$}{Lifetime of dimethyl sulfide}
\nomenclature[A]{$E_{dms}(t)$}{Dimethyl sulfide emissions}
\nomenclature[A]{$\Gamma_{SO_{4}}$}{Sulfate sensitivity to the global mean temperature}
\nomenclature[A]{$\Delta T_{as}(t)$}{Global mean temperature relative to 1850 at time $t$}

Similarly, the concentration of primary organic aerosols (POAs) is defined as:
\begin{align}
  \label{eq:poa}
  C_{POA}(t) = &C_{POA}^{0} + \tau_{OM}^{ind} \alpha_{POM} E_{OC}^{ind}(t) + \tau_{OM}^{lnd} \alpha_{POM} E_{OC}^{lnd}(t) + \nonumber \\
  & \Gamma_{POA} \Delta T_{as}(t)
\end{align}

\nomenclature[A]{$C_{POA}(t)$}{Concentration of primary organic aerosols at time $t$}
\nomenclature[A]{$C_{POA}^{0}$}{Initial concentration of primary organic aerosols}
\nomenclature[A]{$\tau_{OM}^{ind}$}{Lifetime of industrial primary organic aerosols}
\nomenclature[A]{$\alpha_{POM}$}{Conversion of POM from Tg (OC) into Tg (OM)}
\nomenclature[A]{$E_{OC}^{ind}(t)$}{Industrial OC emissions at time $t$}
\nomenclature[A]{$\tau_{OM}^{lnd}$}{Lifetime of land-use primary organic aerosols}
\nomenclature[A]{$E_{OC}^{lnd}(t)$}{Land-use OC emissions at time $t$}
\nomenclature[A]{$\Gamma_{POA}$}{Primary organic aerosol sensitivity to the global mean temperature}

The black carbon (BC) concentration is:

\begin{align}
  \label{eq:bc}
  C_{BC}(t) = &C_{BC}^{0} + \tau_{BC}^{ind} E_{BC}^{ind}(t) + \tau_{BC}^{lnd} E_{BC}^{lnd}(t) + \nonumber \\
  & \Gamma_{BC} \Delta T_{as}(t)
\end{align}

\nomenclature[A]{$C_{BC}(t)$}{Concentration of BC at time $t$}
\nomenclature[A]{$C_{BC}^{0}$}{Initial concentration of BC}
\nomenclature[A]{$\tau_{BC}^{ind}$}{Lifetime of industrial BC}
\nomenclature[A]{$E_{BC}^{ind}(t)$}{Industrial BC emissions at time $t$}
\nomenclature[A]{$\tau_{BC}^{lnd}$}{Lifetime of land-use BC}
\nomenclature[A]{$E_{BC}^{lnd}(t)$}{Land-use BC emissions in time $t$}
\nomenclature[A]{$\Gamma_{BC}$}{BC sensitivity to the global mean temperature}

The concentration of nitrate aerosols is:

\begin{align}
  \label{eq:no3}
  C_{NO_{3}}(t) = &C_{NO_{3}}^{0} + \tau_{NO_{x}} \left(E_{NO_{x}}^{ind}(t) + E_{NO_{x}}^{lnd}(t) \right) + \nonumber \\
   &\tau_{NH_{3}} \left(E_{NH_{3}}^{ind}(t) + E_{NH_{3}}^{lnd}(t) \right) + \Gamma_{NO_{3}} \Delta T_{as}(t)
\end{align}

\nomenclature[A]{$C_{NO_{3}}(t)$}{Concentration of nitrate aerosols at time $t$}
\nomenclature[A]{$C_{NO_{3}}^{0}$}{Initial concentration of nitrate aerosols}
\nomenclature[A]{$\tau_{NO_{x}}$}{Lifetime of $NO_{x}$}
\nomenclature[A]{$E_{NO_{x}}^{ind}(t)$}{Industrial $NO_{x}$ emissions at time $t$}
\nomenclature[A]{$E_{NO_{x}}^{lnd}(t)$}{Land-use $NO_{x}$ emissions at time $t$}
\nomenclature[A]{$\tau_{NH_{3}}$}{Lifetime of $NH_{3}$}
\nomenclature[A]{$E_{NH_{3}}^{ind}(t)$}{Industrial $NH_{3}$ emissions at time $t$}
\nomenclature[A]{$E_{NH_{3}}^{lnd}(t)$}{Land-use $NH_{3}$ emissions at time $t$}
\nomenclature[A]{$\Gamma_{NO_{3}}$}{$NO_{x}$ sensitivity to the global mean temperature}

The concentration of secondary organic aerosols (SOAs) is:

\begin{align}
  \label{eq:soa}
  C_{SOA}(t) = &C_{SOA}^{0} + \tau_{VOC}\left(E_{VOC}^{ind}(t) + E_{VOC}^{lnd}(t) \right) + \tau_{BVOC} E_{BVOC}(t) + \nonumber \\
  & \Gamma_{SOA} \Delta T_{as}(t)
\end{align}

\nomenclature[A]{$C_{SOA}(t)$}{Concentration of SOAs at time $t$}
\nomenclature[A]{$C_{SOA}^{0}$}{Initial concentration of SOAs}
\nomenclature[A]{$\tau_{VOC}$}{Lifetime of nonmethane volatile organic compounds (NMVOCs)}
\nomenclature[A]{$E_{VOC}^{ind}(t)$}{Industrial NMVOC emissions at time $t$}
\nomenclature[A]{$E_{VOC}^{lnd}(t)$}{land-use NMVOC emissions at time $t$}
\nomenclature[A]{$\tau_{BVOC}$}{Lifetime of biogenic NMVOCs}
\nomenclature[A]{$E_{BVOC}(t)$}{Biogenic NMVOC emissions at time $t$}
\nomenclature[A]{$\Gamma_{SOA}$}{NMVOCs sensitivity to the global mean temperature}

Thus, the direct radiative forcing caused by aerosols and pollutants is:

\begin{equation}
  \label{eq:aero}
  f_{aero}(t)= \alpha_{aero}^{rf} \delta C_{aero}(t)
\end{equation}

\nomenclature[A]{$f_{aero}(t)$}{Direct radiative forcing of aerosol $aero$ at time $t$}
\nomenclature[A]{$\alpha_{aero}^{rf}$}{Radiative efficiency of aerosol $aero$}
\nomenclature[A]{$\delta C_{aero}(t)$}{Aerosol $aero$ concentration at time $t$}

\subsection*{Mineral dust aerosols}

The historical radiative forcing from mineral dust aerosols is obtained from MAGICC 6.0 \cite{Meinshausen2011}. The future forcing level is assumed to remain at a constant value of -0.1 Wm\textsuperscript{-2} after 2005.

\begin{equation}
\label{eq:fmd}
f_{mindust}(t)=-0.1
\end{equation}

\nomenclature[A]{$f_{mindust}(t)$}{Radiative forcing from mineral dust}

\subsection*{Cloud effects}

The tropospheric burden of soluble aerosols can be obtained by:

\begin{equation}
\label{eq:csolaero}
C_{solu}(t) = C_{solu}^{0} + \sum_{aero \in SO_{4}, POA,BC,NO_{3},SOA} \alpha_{solu}^{aero} \left(C_{aero}(t) - C_{aero}^{0}\right)
\end{equation}

\nomenclature[A]{$C_{solu}(t)$}{Number concentrations of soluble aerosols at time $t$}
\nomenclature[A]{$C_{solu}^{0}$}{Initial number concentrations of soluble aerosols}
\nomenclature[A]{$\alpha_{solu}^{aero}$}{Soluble fraction for aerosol $aero$}
\nomenclature[A]{$C_{aero}(t)$}{Aerosol concentration at time $t$}
\nomenclature[A]{$C_{aero}^{0}$}{Initial aerosol concentration}

The cloud forcing effects are estimated by:
\begin{equation}
\label{eq:fcloud}
f_{cloud}(t) = f_{BC}(t) \kappa_{adj}^{BC} + \phi_{solu} ln \left(1 +  \frac{\Delta C_{solu}(t)}{C_{solu}^{0}} \right)
\end{equation}
\nomenclature[A]{$f_{cloud}(t)$}{Cloud forcing effects at time $t$}
\nomenclature[A]{$f_{BC}(t)$}{BC radiative forcing at time $t$}
\nomenclature[A]{$\kappa_{adj}^{BC}$}{Adjustment coefficient of the BC radiative forcing to the cloud forcing effect}
\nomenclature[A]{$\phi_{solu}$}{Intensity effect coefficient for soluble aerosols}

\subsection*{Stratospheric ozone}

The equivalent effective stratospheric chlorine (EESC) concentration is calculated as:

\begin{equation}
\label{eq:eesc}
C_{EESC}(t)=a_{EESC}\left(\sum_{Cl}n_{Cl}f_{Cl}C_{hc}(t,Cl)+\alpha_{br}\sum_{Br}n_{Br}f_{Br}C_{hc}(t,Br)\right)
\end{equation}

\nomenclature[O]{$C_{EESC}(t)$}{EESC concentration at time $t$}
\nomenclature[O]{$n_{Cl}$ and $n_{Br}$}{Numbers of chlorine and bromine atoms, respectively}
\nomenclature[O]{$f_{Cl}$ and $f_{Br}$}{Release efficiencies of stratospheric halogens for chlorine and bromine, respectively}
\nomenclature[O]{$C_{hc}(t,Cl)$ and $C_{hc}(t,Br)$}{Gas mixing rates in the stratosphere for chlorine and bromine, respectively}
\nomenclature[O]{$\alpha_{br}$}{Ratio of effectiveness in ozone depletion between bromine and chlorine}
\nomenclature[O]{$a_{EESC}$}{Fractional release factor of EESC}

The concentration of stratospheric ozone is:

\begin{align}
  \label{eq:co3s}
  C_{O3s}(t) = & C_{O3s}^{0} + \xi_{EESC}^{O3s}\left(C_{EESC}(t) - C_{EESC}^{0} \right) + \nonumber \\
  &\xi_{N_{2}O}^{O3s}\left(1 - \frac{C_{EESC}(t) - C_{EESC}^{0}}{C_{EESC}^{X}} \right) \Delta C_{N_{2}O}^{lag}(t) + \Gamma_{O3s} \Delta T_{as}(t)
\end{align}

\nomenclature[O]{$C_{O3s}(t)$}{Stratospheric ozone concentration at time $t$}
\nomenclature[O]{$C_{O3s}^{0}$}{Initial stratospheric ozone concentration}
\nomenclature[O]{$\xi_{EESC}^{O3s}$}{Stratospheric ozone sensitivity to EESC}
\nomenclature[O]{$C_{EESC}^{0}$}{Initial EESC concentration}
\nomenclature[O]{$\xi_{N_{2}O}^{O3s}$}{Stratospheric ozone sensitivity to $N_{2}O$}
\nomenclature[O]{$C_{EESC}^{X}$}{Nonlinear interaction parameter between the chlorine and nitrogen chemistries}
\nomenclature[O]{$\Delta C_{N_{2}O}^{lag}(t)$}{$N_{2}O$ concentration with time lag at time $t$}
\nomenclature[O]{$\Gamma_{O3s}$}{Stratospheric ozone sensitivity to the global mean temperature}

Thus, the forcing effect of the stratospheric ozone burden can be obtained by:

\begin{equation}
\label{eq:fo3s}
f_{O3s}(t)=\alpha_{O3s}^{rf}\left(C_{O3s}(t)-C_{O3s}^{0}\right)
\end{equation}

\nomenclature[O]{$f_{O3s}(t)$}{Forcing effect of the stratospheric ozone burden at time $t$}
\nomenclature[O]{$\alpha_{O3s}$}{Stratospheric ozone radiative efficiency}

\subsection*{Tropospheric ozone}

The tropospheric ozone concentration is estimated to be:

\begin{align}
  \label{eq:co3t}
  C_{O3t}(t) = &C_{O3t}^{0} + \xi_{CH_{4}}^{O3t} ln \left(1+\frac{\Delta C_{CH_{4}}(t)}{C_{CH_{4}}^{0}} \right) + \Gamma_{O3t} \Delta T_{as}(t) + \nonumber \\
  &\sum_{aero \in NO_{x},CO,VOC} \xi_{aero}^{O3t} \left(E_{aero}^{ind}(t) + E_{aero}^{lnd}(t)\right)
\end{align}

\nomenclature[O]{$C_{O3t}(t)$}{Tropospheric ozone concentration at time $t$}
\nomenclature[O]{$C_{O3t}^{0}$}{Initial tropospheric ozone concentration}
\nomenclature[O]{$\xi_{CH_{4}}^{O3t}$}{Tropospheric ozone sensitivity of the $CH_{4}$ effect}
\nomenclature[O]{$\Gamma_{O3t}$}{Tropospheric ozone sensitivity to the global mean temperature}
\nomenclature[O]{$\xi_{aero}^{O3t}$}{Tropospheric ozone sensitivity of aerosol $aero$}

The radiative forcing from the tropospheric ozone is then calculated as:
\begin{equation}
\label{eq:fo3t}
f_{O3t}(t)=\alpha_{O3t}^{rf}\left(C_{O3t}(t)-C_{O3t}^{0}\right)
\end{equation}

\nomenclature[O]{$f_{O3t}(t)$}{Radiative forcing of the tropospheric ozone at time $t$}
\nomenclature[O]{$\alpha_{O3t}$}{Tropospheric ozone radiative efficiency}

\subsection*{Stratospheric water vapor from CH\textsubscript{4} oxidation}

The forcing effect of the stratospheric water vapor from CH\textsubscript{4} oxidation $f_{H_{2}O}(t)$ is calculated by:

\begin{equation}
\label{eq:fh2o}
f_{H_{2}O}(t)=\alpha_{H_{2}O}^{rf} \sqrt{C_{CH_{4}}^{0}} \left(\sqrt{1 + \frac{\Delta C_{CH_{4}}^{lag}(t)}{C_{CH_{4}}^{0}} } - 1 \right)
\end{equation}

\nomenclature[T]{$f_{H_{2}O}(t)$}{Forcing effect of the stratospheric water vapor from CH\textsubscript{4} oxidation at time $t$}
\nomenclature[T]{$\alpha_{H_{2}O}^{rf}$}{Stratospheric water vapor radiative efficiency}
\nomenclature[T]{$\Delta C_{CH_{4}}^{lag}(t)$}{$CH_{4}$ concentration with time lag at time $t$}

\subsection*{Land-use albedo}

The forcing effect from the land-use albedo is estimated according to the annual mean albedo at the biome and regional scales, using the changes in regional land cover as input following the methods described in ref \cite{Gasser2016}.

\begin{equation}
\label{eq:flcc}
f_{LCC}(t)=-\pi_{trans} \phi_{rsds} \sum_{bio} \alpha_{LCC}^{bio} \frac{\Delta A_{LCC}^{bio}(t)}{\Delta A_{Earth}}
\end{equation}

\nomenclature[S]{$f_{LCC}(t)$}{Land-use albedo forcing at time $t$}
\nomenclature[S]{$\pi_{trans}$}{Global shortwave and upward transmittance}
\nomenclature[S]{$\phi_{rsds}$}{Radiative shortwave and downward flux at the surface}
\nomenclature[S]{$\alpha_{LCC}^{bio}$}{Yearly averaged albedo at the biome scale}
\nomenclature[S]{$\Delta A_{LCC}^{bio}(t)$}{Surface area change in the biome at time $t$}
\nomenclature[S]{$\Delta A_{Earth}$}{Surface area of Earth}

\subsection*{BC on snow}

The forcing effect of BC on snow is determined as a linear function of the BC emission level:

\begin{equation}
\label{eq:fbcsnow}
f_{BCSnow}(t)=a_{BC}+ b_{BC} \left(E_{BC}^{ind}(t) + E_{BC}^{lnd}(t)\right)
\end{equation}

\nomenclature[T]{$f_{BCSnow}(t)$}{Forcing effect of BC on snow}
\nomenclature[T]{$a_{BC}$ and $b_{BC}$}{Forcing scaling parameters of BC on snow}

\subsection*{Natural sources}
Regarding the various natural sources, the volcanic and solar forcings are assumed to be the natural forcing inputs for CMIP6.

\begin{equation}
\label{eq:fvolc}
f_{volc}(t)=f_{volc}^{CMIP6}(t)
\end{equation}

\begin{equation}
\label{eq:fsolar}
f_{solar}(t)=f_{solar}^{CMIP6}(t)
\end{equation}

\nomenclature[T]{$f_{volc}(t)$}{Volcanic forcing effects at time $t$}
\nomenclature[T]{$f_{volc}^{CMIP6}(t)$}{Volcanic forcing effects for CMIP6 at time $t$}
\nomenclature[T]{$f_{solar}(t)$}{Solar irradiance forcing effects at time $t$}
\nomenclature[T]{$f_{solar}^{CMIP6}(t)$}{Solar irradiance forcing effects for CMIP6 at time $t$}

\subsection*{Global mean temperature}

The estimation of the global mean temperature is based on the Diffusion Ocean Energy balance CLIMate (DOECLIM) model by using the total radiative forcing as input \cite{Tanaka2007,Wong2017}. Here, we reestimated the climate sensitivity, vertical ocean diffusivity and radiative forcing coefficient for CO\textsubscript{2} doubling based on the CMIP5 outputs related to each available GCM. The detailed descriptions and equations are contained in the references \cite{Tanaka2007,Wong2017}.

\begin{figure}[ht]
  \centering
  \includegraphics[width=\textwidth]{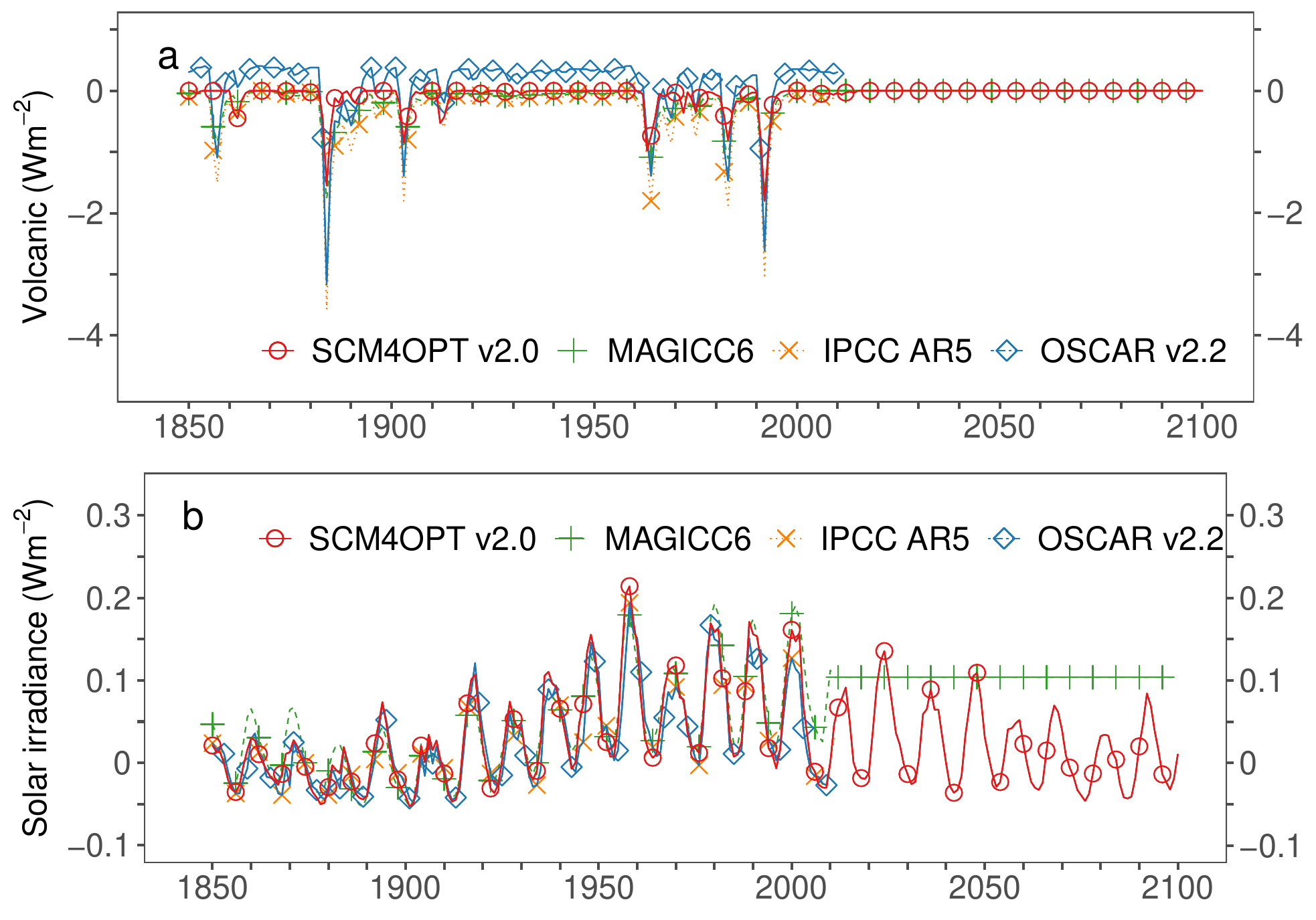}
  \caption{\textbf{Assumptions of the radiative forcings induced by the natural sources of volcanic activity and solar irradiance compared to existing studies.} The volcanic and solar irradiance forcings used in SCM4OPT v2.0 are assumed in accordance with volcanic activity \cite{Zanchettin2016} and solar irradiance \cite{Matthes2017} forcing inputs for CMIP6, and the volcanic forcing is normalized to zero in 1850.}
  \label{fig:rfc_natr}
\end{figure}

\begin{figure}[ht]
  \centering
  \includegraphics[width=\textwidth]{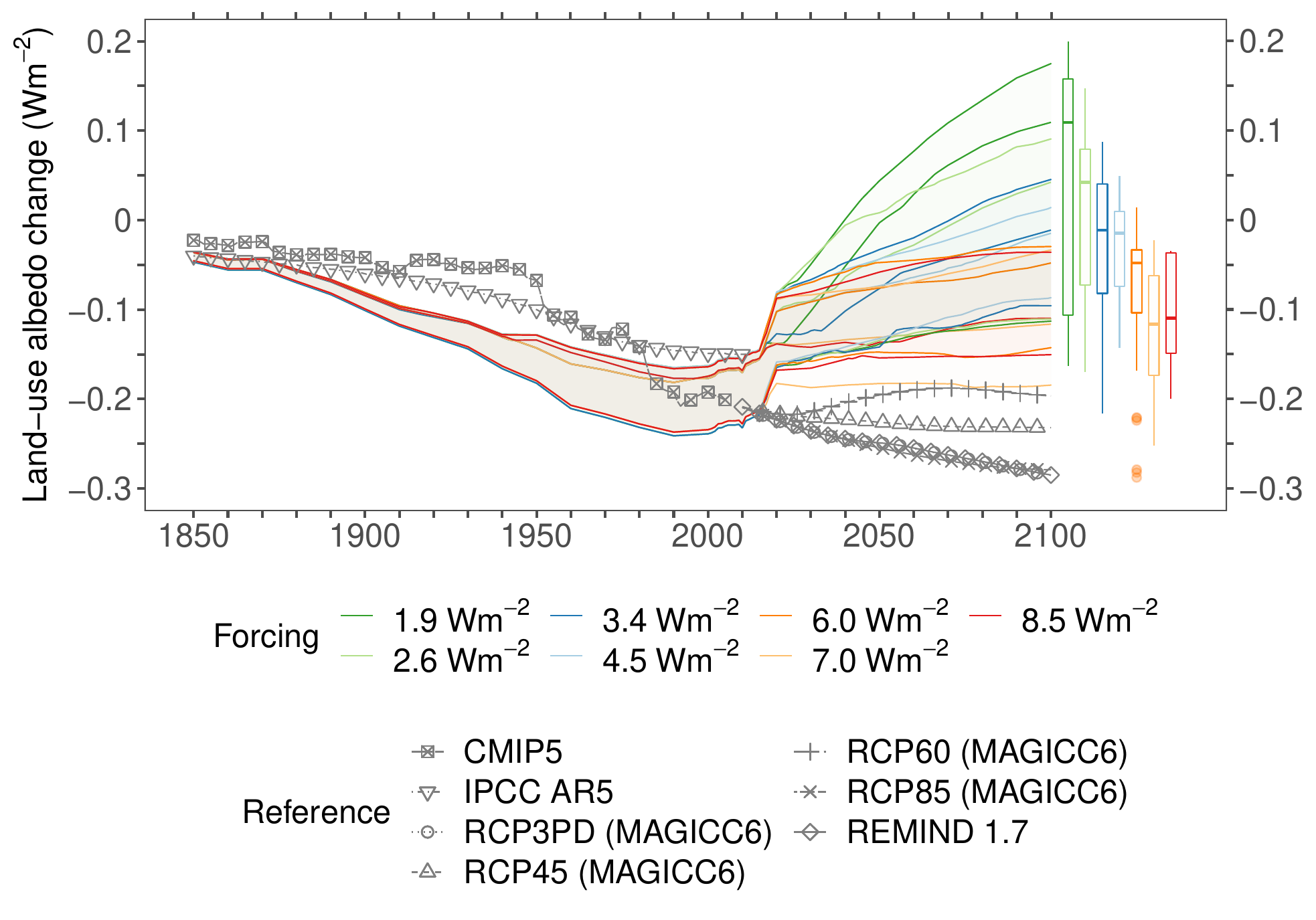}
  \caption{\textbf{Land-use albedo forcings estimated by SCM4OPT v2.0 compared to existing studies.} REMIND 1.7 uses default exogenous values as the future outlook (extracted from the source code, https://www.pik-potsdam.de/research/transformation-pathways/models/remind). The RCP scenarios are produced by MAGICC6\cite{Meinshausen2011b}. The uncertainties in SCM4OPT v2.0 indicate the 17th and 83rd percentiles. The boxplot to the right shows the distributions in 2100, with the upper and lower hinges corresponding to the 25th and 75th percentiles, respectively, where the upper whisker denotes 1.5 times the interquartile range above the 75th percentile, and the lower whisker denotes 1.5 times the interquartile range below the 25th percentile.}
  \label{fig:rfc_LCC}
\end{figure}

\begin{figure}[ht]
  \centering
  \includegraphics[width=\textwidth]{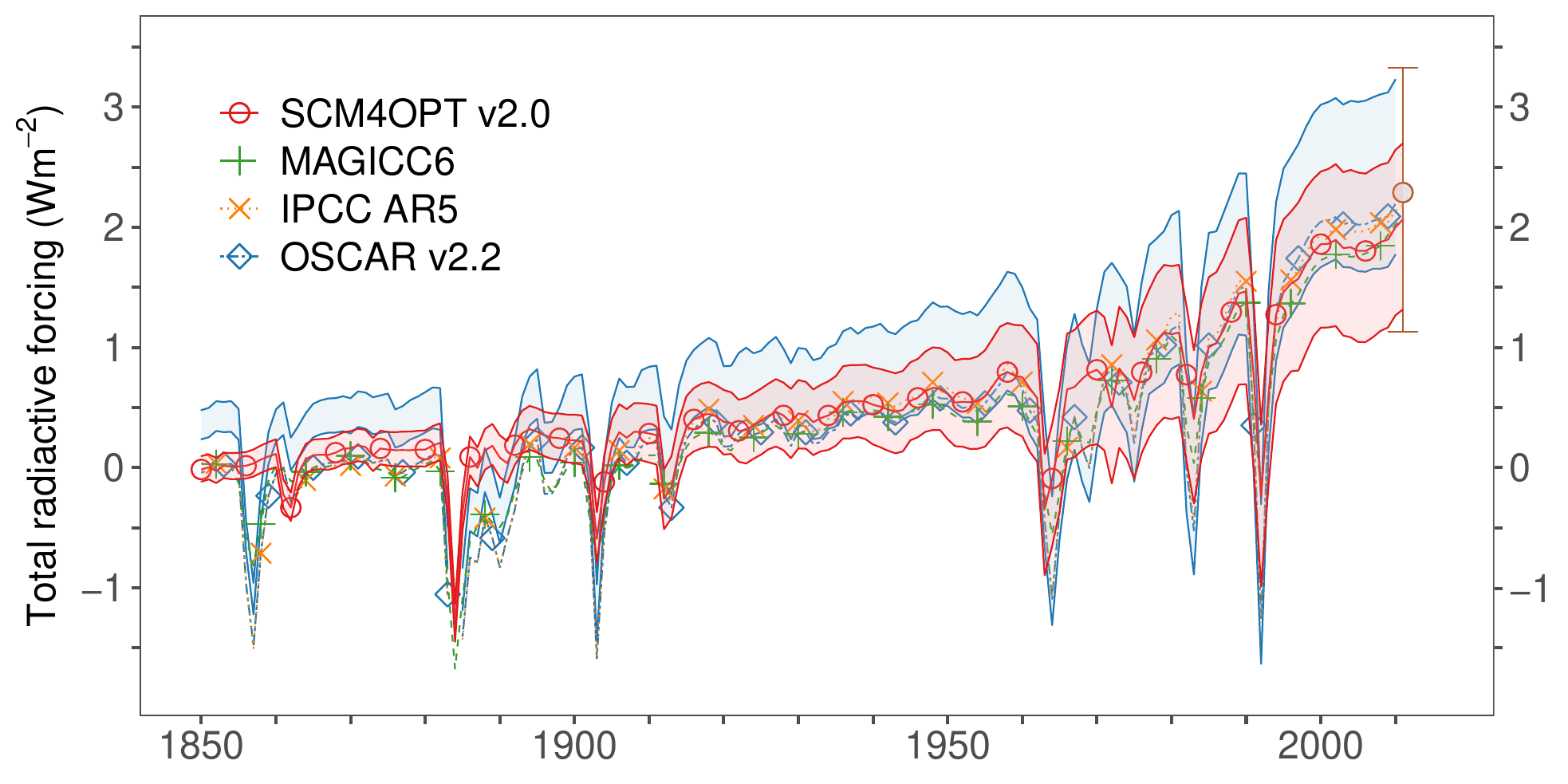}
  \caption{\textbf{Total radiative forcing simulated by SCM4OPT v2.0 compared to existing studies (IPCC AR5\cite{ar5ch8}, MAGICC6\cite{Meinshausen2011} and OSCAR v2.2\cite{Gasser2016}).} The uncertainties in SCM4OPT v2.0 indicate the 17th and 83rd percentiles. The MAGICC6 time series are extracted from RCP calculations\cite{Meinshausen2011b}. The OSCAR v2.2 uncertainties are produced by 500 runs, accounting for the 17th and 83rd percentiles. The error bars in 2011 denote the total anthropogenic radiative forcing relative to 1750 (Figure SPM.5).}
  \label{fig:rfc_tot}
\end{figure}

\begin{figure}[ht]
  \centering
  \includegraphics[width=\textwidth]{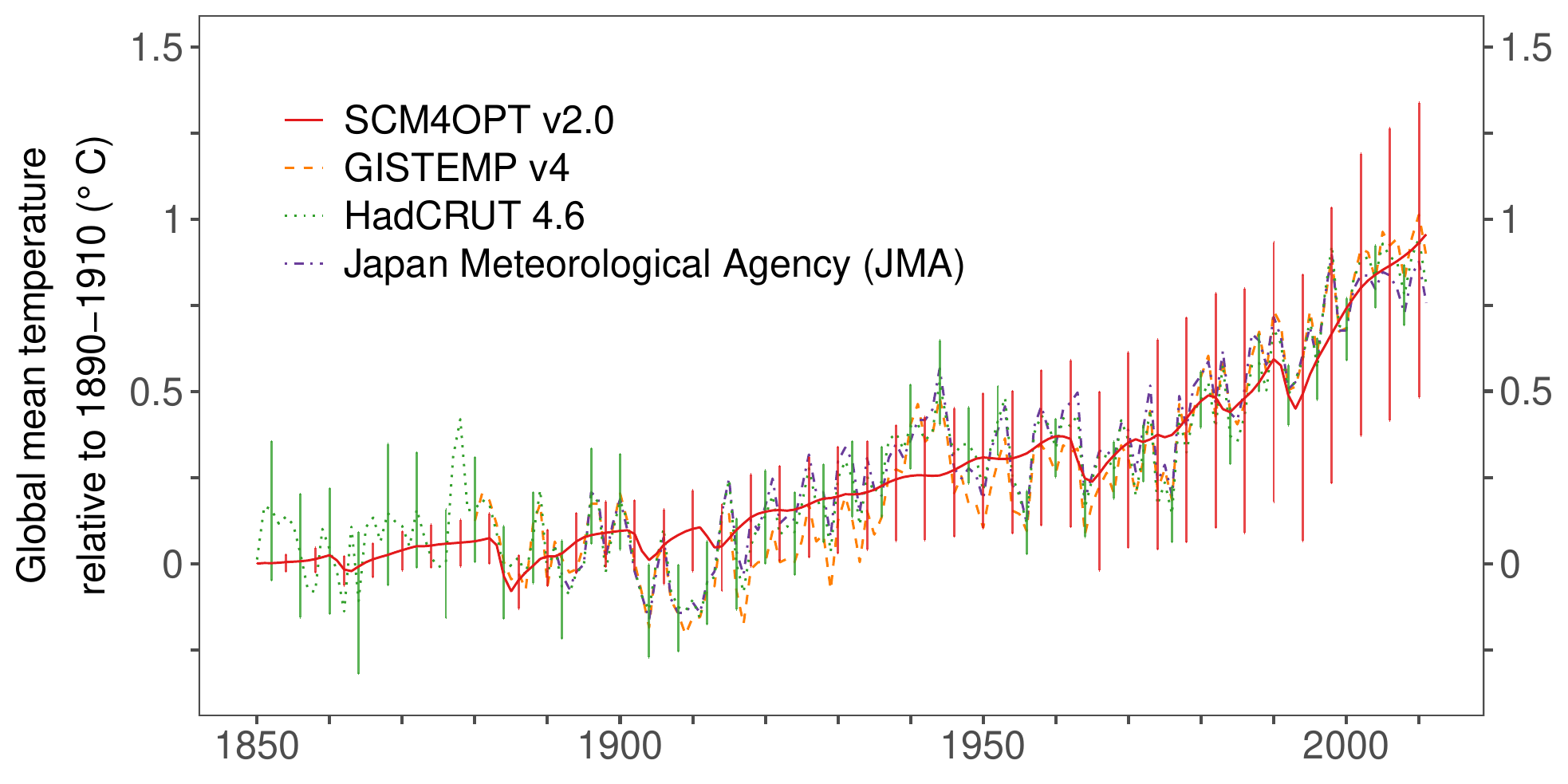}
  \caption{\textbf{Historical global mean temperature increase above the preindustrial level, generated by SCM4OPT v2.0 and compared to existing statistical records.} The anomalies deviate from the average over 1890-1910. The SCM4OPT v2.0 uncertainties result from the emission source- (CEDS\cite{Hoesly2018}, EDGAR v4.3.2\cite{Aardenne2018} and RCP historical\cite{Meinshausen2011b}) and climate uncertainties described in this paper. The uncertainties in HadCRUT 4.6 indicate the 95\% confidence interval of the combined effects of all the uncertainties described in the HadCRUT4 error model. GISTEMP v4 from ref \cite{GISTEMPv4}; HadCRUT 4.6 from ref \cite{Morice2012}; Japan Meteorological Agency (JMA) from ref \cite{JMA2019}.}
  \label{fig:scm_tatm}
\end{figure}

For the simple climate module, the time step was calibrated to be 1/6 year for SCM4OPT v2.0 to avoid possible convergence problems when calculating the ocean carbon cycle \cite{Hartin2015}. The calibrated results are shown in \cref*{fig:rfc_ghg,fig:rfc_aero,fig:rfc_other,fig:rfc_natr,fig:rfc_tot,fig:rfc_LCC,fig:scm_tatm}. We also included the results produced by other models or associated statistical records for comparison purposes.

\printnomenclature

\newpage

\begin{table}
  \small
  \centering
  \caption{Datasets of historical emissions}
  \label{tbl_hist}
  \begin{tabular}{ l l l l l }
    \hline
    Source & Period & Emission & Format & Reference \\
    \hline
    CEDS & 1750-2014 & {\makecell{CO\textsubscript{2}, CH\textsubscript{4}, BC, CO, NH\textsubscript{3}, \\ NMVOC, NO\textsubscript{x}, OC, SO\textsubscript{2}}} & Spatial (sectoral) & {Ref \cite{Hoesly2018}} \\
    EDGAR v4.3.2 & 1970-2012 & {\makecell{CO\textsubscript{2}, CH\textsubscript{4}, N\textsubscript{2}O, BC, CO, NH\textsubscript{3}, \\ NMVOC, NO\textsubscript{x}, OC, SO\textsubscript{2}}} & {\makecell{Regional and sectoral \\ /Spatial (sectoral)}} & {Ref \cite{Aardenne2018}} \\
    EDGAR v4.2 (*) & 1970-2008 & {\makecell{CO\textsubscript{2}, CH\textsubscript{4}, N\textsubscript{2}O, CO, NH\textsubscript{3}, F-gases,\\ NF3, SF6, NMVOC, NO\textsubscript{x}, SO\textsubscript{2}}} & {\makecell{Regional and sectoral \\ /Spatial (sectoral)}} & {Ref \cite{JRCPBL2011}} \\
    PRIMAP v2.0 (**) & 1850-2016 & {\makecell{CO\textsubscript{2}, CH\textsubscript{4}, N\textsubscript{2}O, F-gases, HFCs, \\ PFCs, NF3, SF6}} & Spatial (sectoral) & {Ref \cite{Gutschow2016}} \\
    RCP historical & 1850-2000 & {\makecell{CH\textsubscript{4}, BC, CO, NH\textsubscript{3}, NO\textsubscript{x}, OC, \\ SO\textsubscript{2}, VOC}} & Spatial (sectoral) & {Ref \cite{Lamarque2009}} \\
    \hline
  \end{tabular}
    \begin{tablenotes}
      \item (*) Halogenated gas emissions are used in EDGAR v4.3.2 since these emissions are not included in EDGAR v4.3.2.
      \item (**) N\textsubscript{2}O is employed in the other datasets when not included.
    \end{tablenotes}
\end{table}

\newpage
\normalsize

\begin{table}
  \small
  \centering
  \begin{threeparttable}
  \caption{Datasets of the future scenarios at the various forcing levels}
  \label{tbl_sce}
   \begin{tabular}{ l l l l l }
    \hline
    \makecell{Forcing levels\\ (Wm\textsuperscript{-2})} & Source & Scenario & Reference \\
    \hline
    1.9 & AIM/CGE & {SSP1-1.9, SSP2-1.9}  & {Ref \cite{Fujimori2018}} \\
    1.9 & IAMC    & SSP1-1.9 & {Ref \cite{Gidden2019}} \\
    2.6 & AIM/CGE & {SSP1-2.6, SSP2-2.6, SSP3-2.6(*), SSP4-2.6, SSP5-2.6} & {Ref \cite{Fujimori2018}} \\
    2.6 & IAMC    & SSP1-2.6,{Ref \cite{Gidden2019}} \\
    3.4 & AIM/CGE & {SSP1-3.4, SSP2-3.4, SSP3-3.4, SSP4-3.4, SSP5-3.4} & {Ref \cite{Fujimori2018}} \\
    3.4 & IAMC    & {SSP4-3.4, SSP5-3.4-OS} & {Ref \cite{Gidden2019}} \\
    4.5 & AIM/CGE & {SSP1-4.5, SSP2-4.5, SSP3-4.5, SSP4-4.5, SSP5-4.5} & {Ref \cite{Fujimori2018}} \\
    4.5 & IAMC    & SSP2-4.5 & {Ref \cite{Gidden2019}} \\
    6.0 & AIM/CGE & {SSP1-Baseline, SSP2-6.0, SSP3-6.0, SSP4-Baseline, SSP5-6.0} & {Ref \cite{Fujimori2018}} \\
    6.0 & IAMC    & {SSP3-LowNTCF(**), SSP4-6.0} & {Ref \cite{Gidden2019}} \\
    7.0 & AIM/CGE & {SSP2-Baseline, SSP3-Baseline} & {Ref \cite{Fujimori2018}} \\
    7.0 & IAMC    & SSP3-7.0 & {Ref \cite{Gidden2019}} \\
    8.5 & AIM/CGE & SSP5-Baseline & {Ref \cite{Fujimori2018}} \\
    8.5 & IAMC    & SSP5-8.5 & {Ref \cite{Gidden2019}} \\
    \hline
  \end{tabular}
    \begin{tablenotes}
      \small
      \item (*) The SSP3-2.6 scenario was not available in Table 2 in ref \cite{Fujimori2018}, however, the dataset was provided in https://doi.org/10.7910/DVN/4NVGWA. We retained SSP3-2.6 in our analysis.
      \item (**) The target forcing level of SSP3-LowNTCF was 6.3 Wm\textsuperscript{-2} (Table 1 in ref\cite{Gidden2019}). We classified it to the closest forcing level of 6.0 Wm\textsuperscript{-2}.
    \end{tablenotes}
  \end{threeparttable}
\end{table}

\newpage
\normalsize

\begin{table}
  \centering
  \begin{threeparttable}
  \caption{Datasets of CO\textsubscript{2} emissions from land-use change}
  \label{tbl_luc}
  \begin{tabular}{l l l l}
    \hline
  Source & Period & Format & Reference \\
  \hline
  {Houghton et al. (2012) (*)} & 1960-2010 & Regional & {Ref \cite{Houghton2012,Hansis2015}} \\
  MPIMET& 1850-2005& Spatial grid& {Ref \cite{Raddatz2010}}\\
  PRIMAP v1.2& 1850-2015& Regional& {Ref \cite{Gutschow2016}}\\
  {Smith and Rothwell (2013)}& 1850-2010& Regional& {Ref\cite{Smith2013}}  \\
  \hline
  \end{tabular}
    \begin{tablenotes}
      \small
      \item (*) An updated version \cite{Hansis2015} was used, downloaded from http://www.globalcarbonatlas.org/en/CO2-emissions.
 
    \end{tablenotes}
  \end{threeparttable}
\end{table}

\newpage

{\linespread{1.3}
\begin{table}
  \caption{Please refer to the spreadsheet in the supplementary tables. Atmospheric drivers and radiative forcings. Note: This table is compiled based on Figure SPM.5 in IPCC (2013) and references Gasser et al. (2016) and Su et al. (2017). All emissions from international shipping activities are regional nonattributable.}
  \label{tbl_region}
\end{table}
}

{\linespread{1.3}
\begin{table}
  \caption{Please refer to the spreadsheet in the supplementary tables, Mapping of the eleven regions. Note: The spatial mapping is based on Natural Earth data (https://www.naturalearthdata.com), and 1:10 m cultural vectors are applied. Columns 2 and 3 are extracted from Natural Earth maps. ADM0\_A3 are the alpha-3 codes defined for each country or region.}
  \label{tbl_sector}
\end{table}
}

{\linespread{1.3}

\begin{table}
  \centering
  \begin{threeparttable}
 
  \footnotesize
  \caption{Equilibrium climate sensitivity (ECS) used in this study compared to other references}
  \label{tbl_clims}
  \begin{tabular}{l r r r r r r r r}
    \hline
    Model & This study &	Ref\cite{Andrews2012} &	Ref\cite{Forster2013}	& Ref\cite{IPCC2014} & Ref\cite{Sherwood2014} & Ref\cite{Gregory20140417} & Ref\cite{Tsutsui2017} &	Ref\cite{Mauritzen2017}\\
    \hline
    ACCESS1-0&3.88& -&3.83&3.8&3.79&3.45&3.76&3.8\\
    ACCESS1-3&3.59& -& -& -&3.45&2.8&3.22& -\\
    bcc-csm1-1&2.80& -&2.82&2.8&2.88& -&2.73&2.8\\
    bcc-csm1-1-m&2.79& -&2.87&2.9& -& -&3.1& -\\
    BNU-ESM&4.11(*)& -& -&4.1&4.11& -&4.08&4.1\\
    CanESM2&3.66&3.69&3.69&3.7&3.68&3.6&3.63&3.7\\
    CCSM4&2.90& -&2.89&2.9&2.92& -&2.8&2.9\\
    CNRM-CM5&3.27&3.25&3.25&3.3&3.25&3.16&3.07&3.3\\
    CNRM-CM5-2&3.46& -& -& -& -& -& -& -\\
    CSIRO-Mk3-6-0&4.24&4.08&4.08&4.1&3.99&2.96&3.55&4.1\\
    FGOALS-g2&3.45(*)& -& -& -&3.45& -&2.46&3.45\\
    FGOALS-s2&4.16(*)& -&4.17& -&4.16& -&4.14&4.16\\
    GFDL-CM3&3.97&3.97&3.97&4&3.96&3.2&3.85&4\\
    GFDL-ESM2G&2.57&2.39&2.39&2.4&2.38& -&1.81& -\\
    GFDL-ESM2M&2.71&2.44&2.44&2.4&2.41& -&2.23&2.4\\
    HadGEM2-ES&4.58&4.59&4.59&4.6&4.55&4.32&4.6&4.6\\
    IPSL-CM5A-LR&4.05&4.13&4.13&4.1&4.1&3.46&3.92&4.1\\
    IPSL-CM5A-MR&4.11& -& -& -& -&3.4& -& -\\
    IPSL-CM5B-LR&2.64& -&2.61&2.6&2.59& -&2.43&2.6\\
    MIROC5&2.70&2.72&2.72&2.7&2.71&2.12&2.22&2.7\\
    MIROC-ESM&4.67&4.67&4.67&4.7&4.65&3.47&3.88&4.7\\
    MPI-ESM-LR&3.64&3.63&3.63&3.6&3.6&3.08&3.27& -\\
    MPI-ESM-MR&3.48& -& -& -&3.44&2.94&3.14&3.4\\
    MPI-ESM-P&3.47&3.45&3.45& -&3.42& -&3.07& -\\
    MRI-CGCM3&2.60&2.6&2.6&2.6&2.59&2.19&2.52&2.6\\
    NorESM1-M&2.82&2.8&2.8&2.8&2.83&2.11&2.48&2.8  \\
    \hline
 
  \end{tabular}
    \begin{tablenotes}
      \small
      \item (*) The ECS values for BNU-ESM, FGOALS-g2 and FGOALS-s2 are retrieved from ref\cite{Sherwood2014}. All other values in this study are estimated by using the standard regression method \cite{Gregory2004,Forster2013} based on the available CMIP5 experiments of the preindustrial control (piControl) and abrupt 4xCO\textsubscript{2} scenario (abrupt4xCO2). (**) Based on Table 9.5 in IPCC AR5-WG1 \cite{IPCC2014}.
    \end{tablenotes}
 
  \end{threeparttable}

  \end{table}

}

{\linespread{1.3}
\begin{table}
  \caption{Please refer to the spreadsheet in the supplementary tables, Sector mapping. Note: (*) The AIM/CGE negative CO2 and land-use CO2 emissions are extracted from the regional dataset rather than from the spatial dataset. (**) Forest burning and grassland burning levels are adjusted based on the percentage share in 2012 in EDGAR v4.3.2.}
  \label{tbl_driver}
\end{table}
}

\begin{table}
  \small
  \centering
  \caption{An overview of the iterations regarding the climate system, scenarios, regions, sectors and emissions}
  \label{tbl_unc}
  \begin{tabular}{ l l l }
    \hline
    Sources & Quantity & Composition \\
    \hline
    Climate system & 63 & {\makecell[l]{Terrestrial carbon cycle, ocean carbon cycle, aerosols and \\ pollutants, climate influences, cloud effects, climate system}} \\
    Scenarios & 7 & {\makecell[l]{1.9 Wm\textsuperscript{-2}, 2.6 Wm\textsuperscript{-2}, 3.4 Wm\textsuperscript{-2}, 4.5 Wm\textsuperscript{-2}, 6.0 Wm\textsuperscript{-2}, 7.0 Wm\textsuperscript{-2} \\ and 8.5 Wm\textsuperscript{-2}}} \\
    Regions & 11 & {\makecell[l]{CHN, IND, JPN, RUS, USA, AFR, EUR, LAM, MEA, OAS, ROW}} \\
    Sectors & 12 & {\makecell[l]{Agriculture, agricultural waste burning, domestic housing \\ and commercial, energy, industry, industrial solvents, surface \\ transportation, waste treatment, open forest burning, open grassland \\ burning, aviation and international shipping}}\\
    Emissions & 48 & {\makecell[l]{Industrial CO\textsubscript{2}, land-use CO\textsubscript{2}, CH\textsubscript{4}, N\textsubscript{2}O, BC, CO, NH\textsubscript{3}, NO\textsubscript{x}, OC,\\ SO\textsubscript{2}, VOCs and halogenated gases (a total of 37 gases including \\HFC-23, HFC-32, HFC-125, HFC-134a, HFC-143a, HFC-152a, \\HFC-227ea, HFC-236fa, HFC-245fa, HFC-365mfc, HFC-43-10mee,\\ CF\textsubscript{4}, C\textsubscript{2}F\textsubscript{6}, C\textsubscript{3}F\textsubscript{8}, c-C\textsubscript{4}F\textsubscript{8}, C\textsubscript{4}F\textsubscript{10}, C\textsubscript{5}F\textsubscript{12}, C\textsubscript{6}F\textsubscript{14}, C\textsubscript{7}F\textsubscript{16}; SF\textsubscript{6}, NF\textsubscript{3}, \\CFC-11, CFC-12, CFC-113, CFC-114, CFC-115, CCl\textsubscript{4}, CH\textsubscript{3}CCl\textsubscript{3}, \\HCFC-22, HCFC-141b, HCFC-142b, Halon-1211, Halon-1202, \\Halon-1301, Halon-2402, CH\textsubscript{3}Br and CH\textsubscript{3}Cl)}} \\
    \hline
  \end{tabular}
    % \begin{tablenotes}
    % \end{tablenotes}
\end{table}

\clearpage
\newpage

{\linespread{1.3}
\begin{figure}[ht]
  \includegraphics[width=0.9\textwidth]{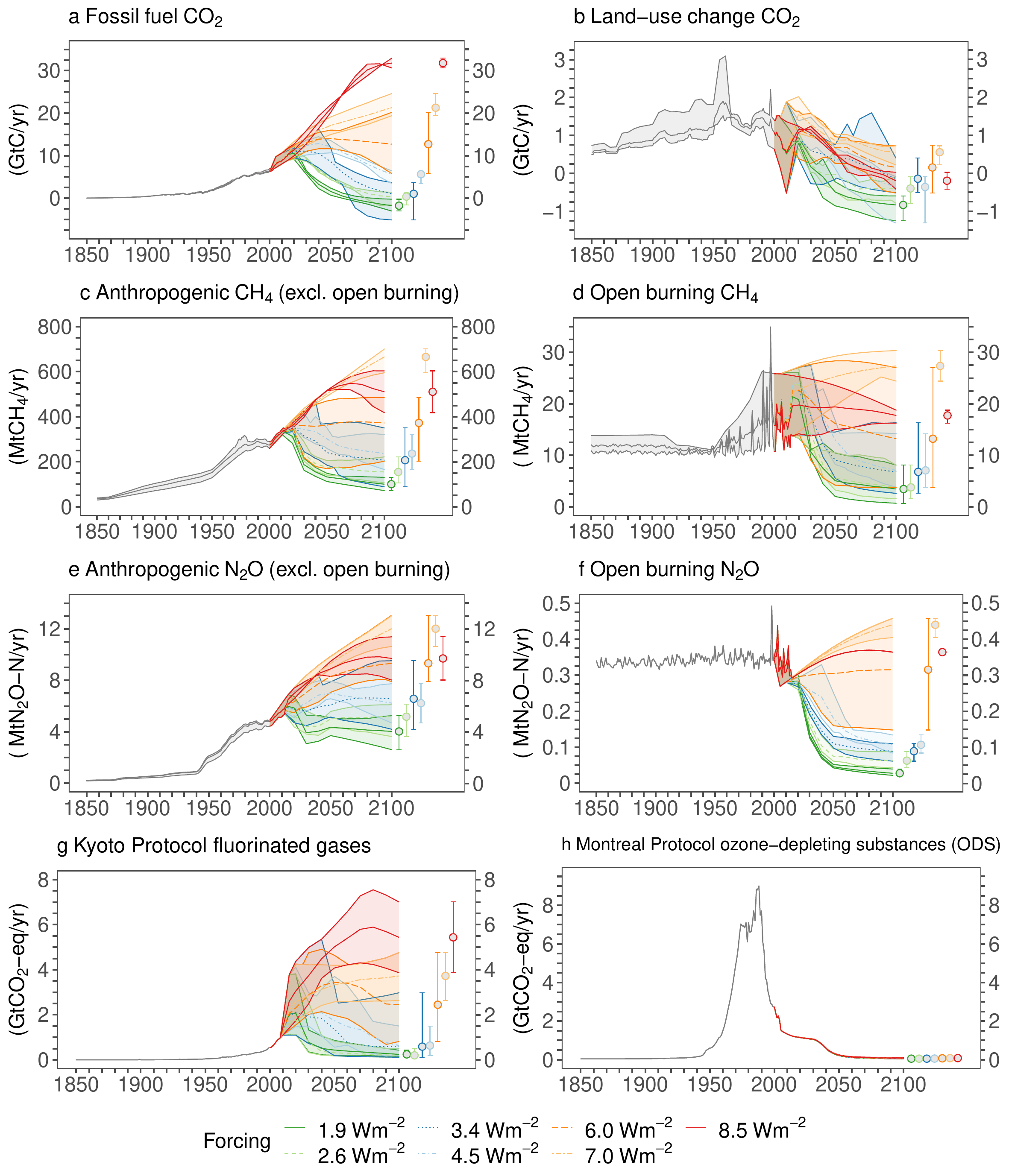}
  \centering
  \caption{\small{\textbf{Historical and future GHG emissions.} The future projections include seven forcing levels, namely, 1.9 Wm\textsuperscript{-2}, 2.6 Wm\textsuperscript{-2}, 3.4 Wm\textsuperscript{-2}, 4.5 Wm\textsuperscript{-2}, 6.0 Wm\textsuperscript{-2}, 7.0 Wm\textsuperscript{-2} and 8.5 Wm\textsuperscript{-2}. The uncertainty ranges denote the upper and lower trends. The error bars to the right show the upper and lower trends in 2100 at each forcing level. Open burning includes the emissions from agricultural waste burning, forest fires and grassland fires. Sources: the historical emissions stem from ref \cite{Lamarque2009,Gutschow2016,Aardenne2018,Hoesly2018}; the future trends stem come ref \cite{Fujimori2018,Gidden2019}; land-use CO\textsubscript{2} originates from ref \cite{Raddatz2010,Smith2013,Gutschow2016,gcp2018}; open burning is from ref \cite{Marle2017}.}}
  \label{fig:ghg}
\end{figure}
}

\newpage

{\linespread{1.3}
\begin{figure}[ht]
  \includegraphics[width=0.8\textwidth]{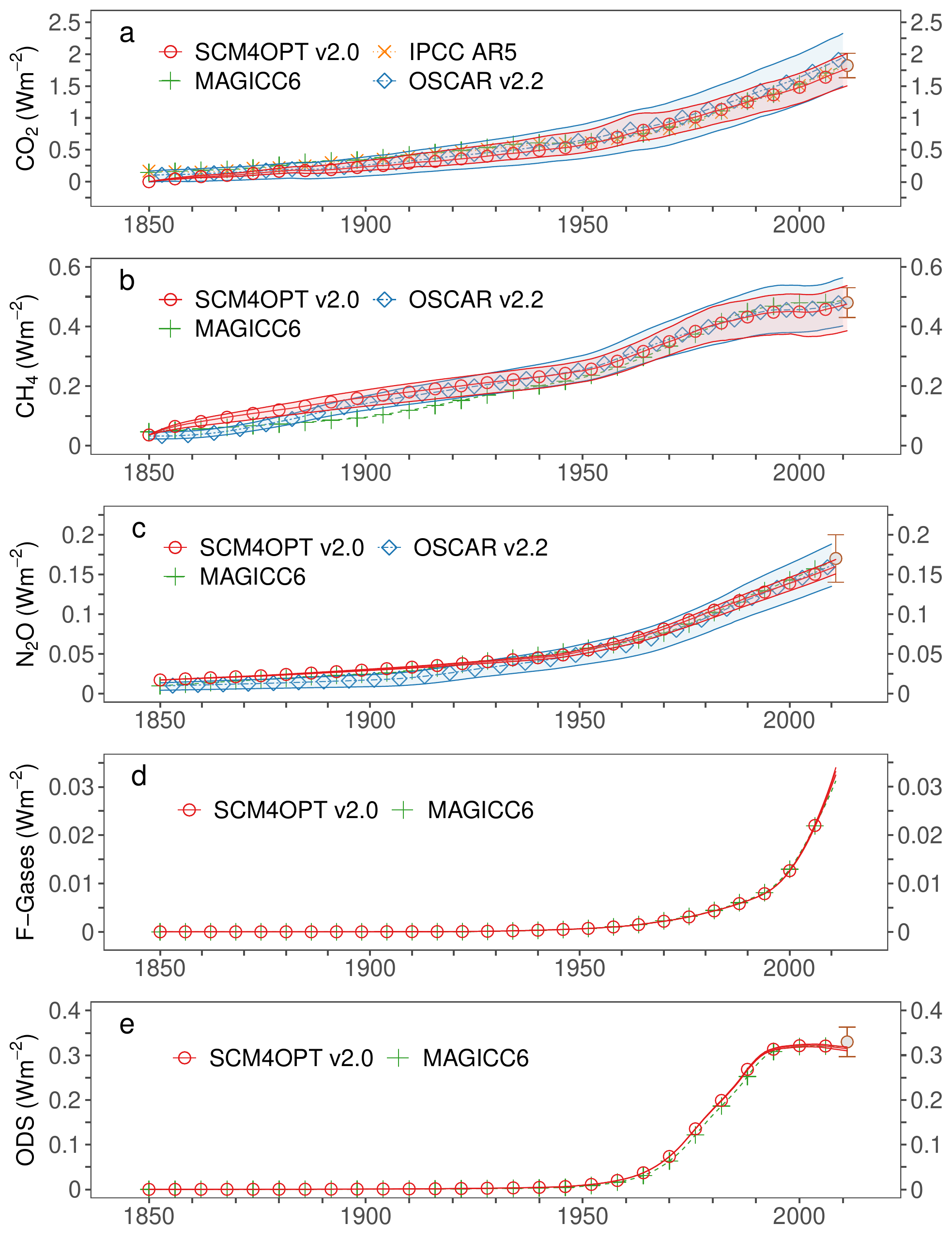}
  \centering
  \caption{\small{\textbf{Simulation of the radiative forcings induced by greenhouse gases (GHGs) compared to existing studies (IPCC AR5\cite{ar5ch8}, MAGICC6\cite{Meinshausen2011} and OSCAR v2.2\cite{Gasser2016}).} The uncertainties in SCM4OPT v2.0 indicate the 17th and 83rd percentiles. The MAGICC6 time series are extracted from RCP calculations \cite{Meinshausen2011b}. The OSCAR v2.2 uncertainties are produced by 500 runs, accounting for the 17th and 83rd percentiles, downloaded from https://github.com/tgasser/OSCARv2. The error bars in 2011 denote the forcing values over the period of 1750-2011 in IPCC AR5 (Table 8.2).}}
  \label{fig:rfc_ghg}
\end{figure}
}

{\linespread{1.3}
\begin{figure}[ht]
  \centering
  \includegraphics[width=0.8\textwidth]{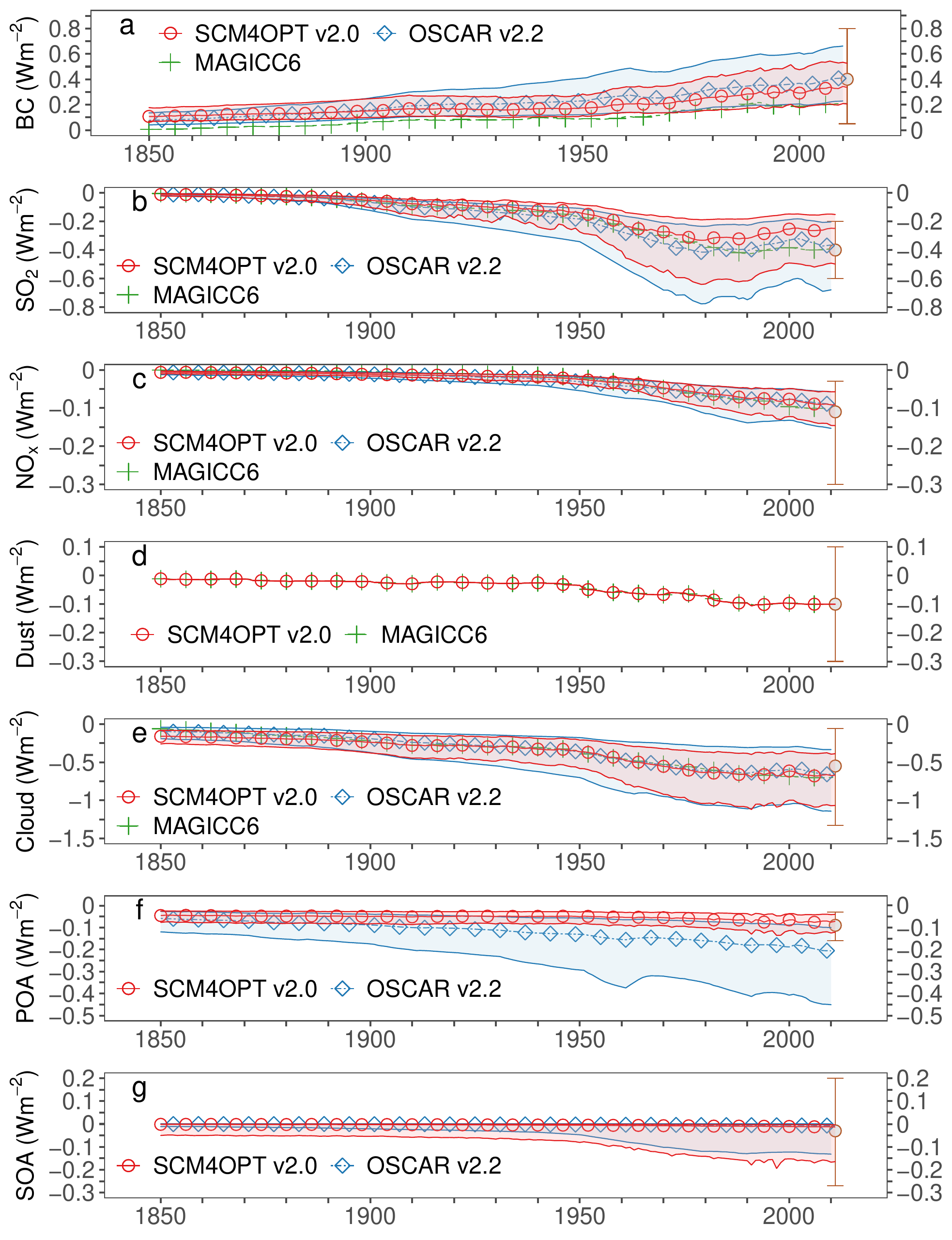}
  \caption{\small{\textbf{Simulation of the radiative forcings induced by aerosols and pollutants, compared to existing studies (IPCC AR5\cite{ar5ch8,ar5wg1spm}, MAGICC6\cite{Meinshausen2011} and OSCAR v2.2\cite{Gasser2016}).} The uncertainties in SCM4OPT v2.0 indicate the 17th and 83rd percentiles. The MAGICC6 time series are extracted from RCP calculations \cite{Meinshausen2011b}. The OSCAR v2.2 uncertainties are produced by 500 runs, accounting for the 17th and 83rd percentiles. The error bars in 2011 denote the forcing values over the period of 1750-2011 in IPCC AR5 (Table 8.4 and Figure SPM.5).}}
  \label{fig:rfc_aero}
\end{figure}
}

{\linespread{1.3}
\begin{figure}[ht]
  \centering
  \includegraphics[width=0.8\textwidth]{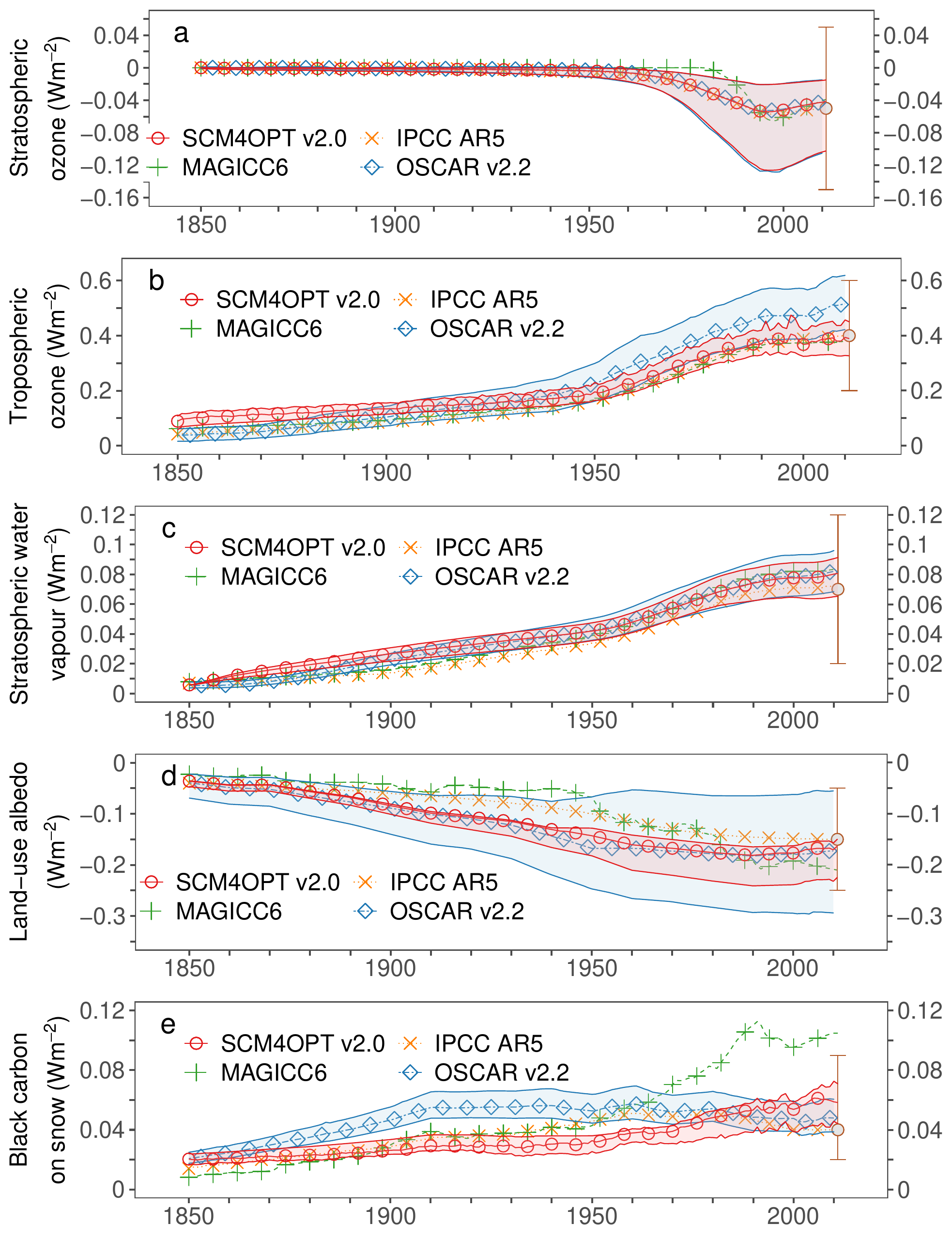}
  \caption{\small{\textbf{Simulation of the radiative forcings induced by human activities, other than the GHGs and aerosols and pollutants above, compared to existing studies (IPCC AR5\cite{ar5ch8}, MAGICC6\cite{Meinshausen2011} and OSCAR v2.2\cite{Gasser2016}).} The uncertainties in SCM4OPT v2.0 indicate the 17th and 83rd percentiles. The MAGICC6 time series are extracted from RCP calculations\cite{Meinshausen2011b}. The OSCAR v2.2 uncertainties are produced by 500 runs, accounting for the 17th and 83rd percentiles. The error bars in 2011 denote the forcing values over the period of 1750-2011 in IPCC AR5 (Table 8.6).}}
  \label{fig:rfc_other}
\end{figure}
}

{\linespread{1}
\begin{figure}[ht]
  \includegraphics[width=0.9\textwidth]{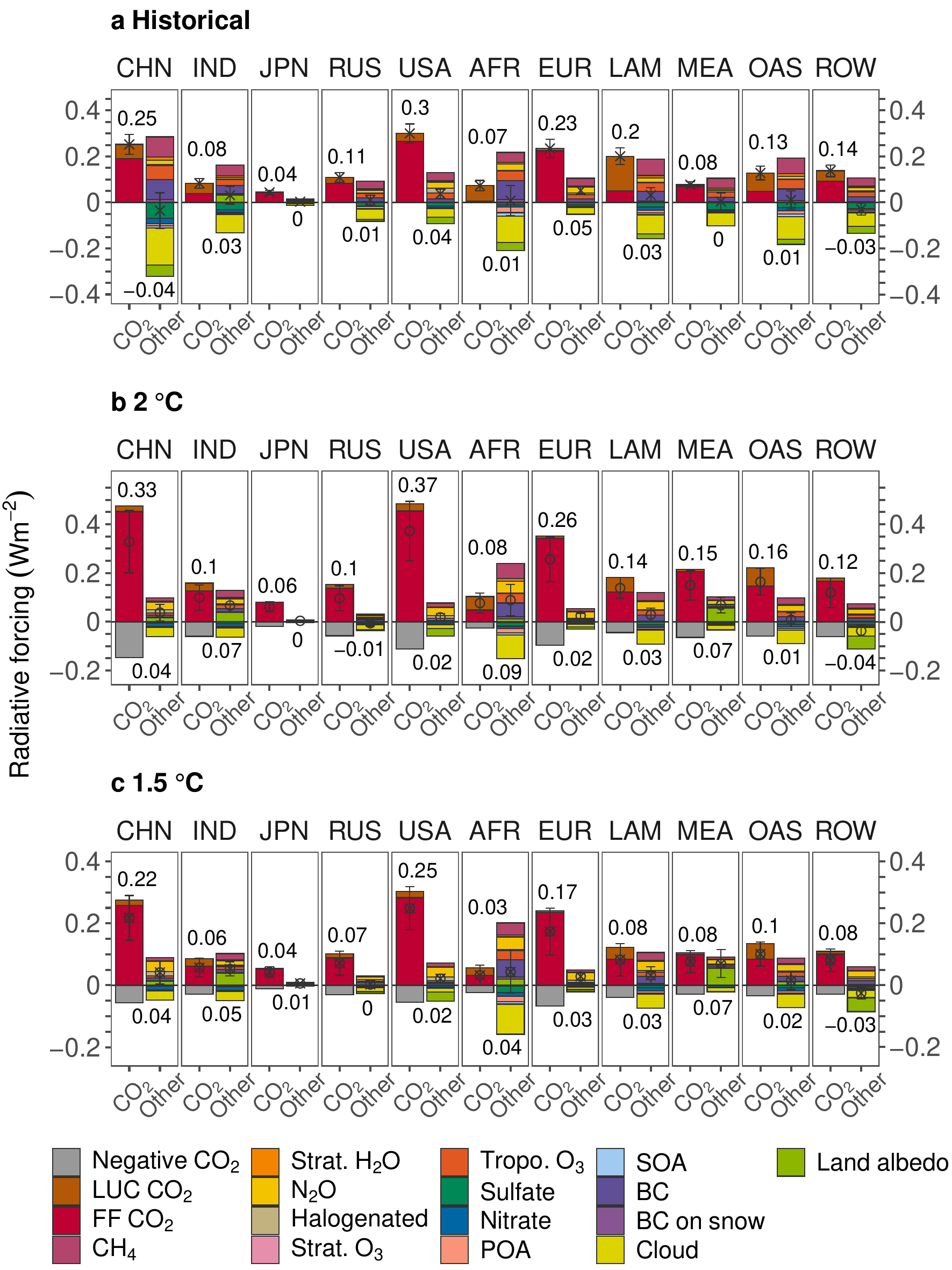}
  \centering
  \caption{\small{\textbf{Regional forcings are decomposed into CO\textsubscript{2}-induced forcings and those not directly related to CO\textsubscript{2}.} a, Historical period (1850-2016); b, 2 \textdegree C (1850-2100); c, 1.5 \textdegree C (1850-2100). The direct CO\textsubscript{2} emissions are separated into fossil-fuel CO\textsubscript{2} (FF CO\textsubscript{2}), land-use CO\textsubscript{2} (LUC CO\textsubscript{2}), and negative CO\textsubscript{2} emissions, if applicable. The value on top of the bar indicates the mean value summing all components of the left CO\textsubscript{2} bar. The value at the bottom of the bar indicates the mean value summing all components of the right bar. All uncertainties are represented as one standard deviation.}}
  \label{fig:reg_grp}
\end{figure}
}

\clearpage

\bibliographystyle{iopart-num}
\bibliography{contrib}

\end{document}